\documentclass[
  aps,
  prd,
  reprint,
  superscriptaddress,
  nofootinbib,
  floatfix,
  longbibliography
]{revtex4-2}

\usepackage{amsmath,amssymb,bm,mathrsfs}
\usepackage{graphicx}
\usepackage{slashed}
\usepackage{booktabs}
\usepackage{xcolor}
\usepackage[normalem]{ulem}
\usepackage{hyperref}
\usepackage{tikz}
\usepackage{placeins}
\usetikzlibrary{arrows.meta}
\graphicspath{{upload/}{figures/}{./}}

\hypersetup{
  colorlinks=true,
  citecolor=blue,
  linkcolor=blue,
  urlcolor=blue
}

\newcommand{\ii}{\mathrm{i}}
\newcommand{\dd}{\mathrm{d}}
\newcommand{\Tr}{\operatorname{Tr}}
\newcommand{\sgn}{\operatorname{sgn}}

\newcommand{\Ree}{\operatorname{Re}}

\begin{document}

\title{Neutral and Charged Pion Modes in Magnetized Quark Matter with Chiral Imbalance}

\author{Shijie Zhai}
\email{zhaishijie24@mails.ucas.ac.cn}
\affiliation{School of Nuclear Science and Technology, University of Chinese Academy of Sciences, Beijing 101408, China}

\author{Lang Yu}
\email{yulang@jlu.edu.cn}
\affiliation{College of Physics, Jilin University, Changchun 130012, China}

\author{Jie Mei}
\email{meijie@ucas.ac.cn}
\affiliation{School of Nuclear Science and Technology, University of Chinese Academy of Sciences, Beijing 101408, China}

\author{Mei Huang}
\email{huangmei@ucas.ac.cn}
\affiliation{School of Nuclear Science and Technology, University of Chinese Academy of Sciences, Beijing 101408, China}

\date{\today}

\begin{abstract}

We investigate neutral and charged pion correlations
in a two-flavor Nambu--Jona-Lasinio (NJL) model
at finite temperature in the presence of a uniform
magnetic field and a prescribed chiral chemical
potential $\mu_5$, using the mean-field approximation
and the single-channel random phase approximation (RPA).
The pion masses are defined by the selected
positive-energy zeros of the real parts of the
corresponding inverse propagators.
We demonstrate that, at zero external spatial momentum,
the complete Dirac trace in the neutral channel
eliminates mixed-branch contributions, thereby
precluding a finite-energy Landau cut.
In contrast, the charged channel allows
non-diagonal Landau-level transitions and exhibits
both pair-production and thermal-scattering contributions.
With increasing $\mu_5$, the neutral-pion mass
exhibits only a weak variation at low temperature,
while the rapid decrease of the mass shifts toward
higher temperatures.
The charged-pion mass decreases with increasing
$\mu_5$ over the selected solution segments.
This behavior may help account for the lattice QCD results.
At zero temperature and fixed $\mu_5$, increasing
the magnetic field lowers the neutral-pion mass
but raises the charged-pion mass.

\end{abstract}

\maketitle

\section{Introduction}
\label{sec:introduction}

Strong magnetic fields can be generated in noncentral
relativistic heavy-ion collisions by spectator protons.
Although initially intense, these fields are short-lived,
and their evolution depends on the collision dynamics
and the response of the produced medium
\cite{Skokov2009Add,Voronyuk2011Add,
DengHuang2012Add,Tuchin2016Add,Huang:2015oca}.
They have motivated extensive studies of anomalous
transport phenomena, such as the chiral magnetic effect(CME),
as well as the influence of magnetic fields on the
QCD phase structure
\cite{Kharzeev2008,Fukushima2008,
MooreTassler2011Add,LiWang2020}.
In particular, magnetic fields modify the quark spectrum
and chiral dynamics, leading to magnetic catalysis(MC)
at low temperatures and inverse magnetic catalysis(IMC)
near the chiral crossover
\cite{Gusynin1996,Miransky2015,
Bali2012a,Bali2012b,Chao:2013qpa,
BruckmannEndrodiKovacs2013,
MuellerPawlowski2015,
AndersenNaylorTranberg2016}.

Pions provide important probes of the magnetic response
of low-energy QCD, as they are pseudo-Goldstone bosons
associated with spontaneous chiral symmetry breaking.
A magnetic field breaks isospin symmetry and leads to
different responses of the neutral and charged pion
channels.
Consequently, their masses can exhibit distinct
magnetic-field dependences.
These effects have been investigated in NJL models
and other effective approaches
\cite{Coppola2018,Coppola2019,GomezDumm:2023owj,
Liu:2018zag,Fayazbakhsh:2012vr,
Avancini:2018svs,Avancini2017,
Andersen2012Add,HattoriYamamoto2019,
Wang:2017vtn,Avancini:2015ady,
LiCaoHe2021,Avancini:2021pmi,
Xing:2021kbw,Coppola:2024mixing}.
Finite-temperature studies have further examined
pion correlations and their spectral properties
in magnetic fields
\cite{Mao2019,Sheng2021,LiMao2023Mott,Mei2026,
Sheng:2021evj,Mei:2022dkd}.

Zero-temperature lattice QCD calculations provide
nonperturbative benchmarks for the magnetic response
of pion properties
\cite{Bali2018Add,Ding2021,Ding2026,Ding:2020jui}.
They show that the connected neutral-pion mass
decreases with increasing magnetic field,
whereas the charged-pion ground-state energy
exhibits a nonmonotonic field dependence, where it increases with magnetic field at first and starts decreasing and then saturates for larger magnetic field.
Understanding these distinct responses, particularly
the nonmonotonic behavior of the charged-pion
ground-state energy, remains an important challenge
for effective model studies.
Although various theoretical approaches have
investigated pion properties in magnetic fields,
a satisfactory quark-level explanation of the
lattice results, especially the charged-pion
magnetic-field dependence, is still desirable
\cite{Hattori2016,Andreichikov2017,Wen:2023qcz,
Xu:2020yag,Lin:2022ied,WangMixing2026Add}.

Chiral imbalance provides an additional aspect
for studying the response of strongly interacting
matter to magnetic fields.
It is associated with axial-charge production through
topological transitions and has been widely discussed
in connection with the chiral magnetic effect
\cite{MooreTassler2011Add,Kharzeev2008,
Fukushima2008,LiWang2020}.
In effective descriptions, chiral imbalance is commonly
parameterized by a chiral chemical potential $\mu_5$.
Since the axial charge is not exactly conserved in QCD,
we treat $\mu_5$ as an externally prescribed axial source,
without modeling its generation or relaxation
\cite{Astrakhantsev2021,Ruggieri2020,Espriu2020}.
Previous lattice and effective-model studies have
investigated the response of chiral symmetry breaking
and restoration to this source
\cite{Yamamoto2011Add,Braguta2016Add,Yu2016,
RuggieriPeng2016Add}.

Meson correlations have been studied separately
in magnetic fields and in chirally imbalanced matter.
In particular, Ref.~\cite{Mei2026} investigated
neutral and charged pion spectral functions in a
magnetic field at finite temperature and vanishing
$\mu_5$, while Ref.~\cite{Ghosh2024} studied scalar
and pseudoscalar correlations at finite $\mu_5$
and $B=0$.
Magnetized quark matter with chiral imbalance has
also been considered in related model studies
\cite{Chaudhuri2022}.
A systematic study combining these two backgrounds
is therefore useful for examining the magnetic
responses of both pion channels and their dependence
on chiral imbalance.

Motivated by the lattice QCD results, we investigate
neutral and charged pion correlations in the
two-flavor NJL model at finite temperature,
in the simultaneous presence of a magnetic field
and an external axial source $\mu_5$.
Within the mean-field approximation and the
single-channel random phase approximation,
we investigate the dependence of the pion masses
on temperature, magnetic field, and chiral imbalance.
At zero external spatial momentum, we find that
the complete Dirac trace in the neutral-pion channel
eliminates mixed-branch contributions, even at
finite $\mu_5$.
In contrast, the charged-pion mass decreases
with increasing $\mu_5$ over the selected
solution segments.
These results provide a quark-level description
of the different magnetic responses of the two
pion channels and offer theoretical insight into
the phenomena observed in lattice QCD.

The remainder of this paper is organized as follows.
Section~\ref{sec:model} introduces the model and
the quark spectrum.
Sections~\ref{sec:pi0} and \ref{sec:picharged}
present the analytical properties and numerical results
for the neutral and charged pion channels, respectively.
Section~\ref{sec:conclusions} summarizes our main
conclusions.
The explicit analytical reduction of the neutral-pion
polarization function is presented in Appendix~A,
while the residue kernels for the charged pion
are collected in Appendix~B.

\section{Model}
\label{sec:model}

We consider the two-flavor NJL model with equal
current quark masses \cite{Nambu1961a,Nambu1961b,Klevansky1992,Hatsuda1994,Buballa2005},
\begin{equation}
  \begin{aligned}
  \mathcal L
  ={}&
  \bar\psi
  \left(
    \ii\slashed D-m_0+\mu _5\gamma^0\gamma^5
  \right)\psi
  \\
  &+G_S\left[
    (\bar\psi\psi)^2+
    (\bar\psi\ii\gamma^5\bm\tau\psi)^2
  \right],
  \end{aligned}
  \label{eq:lagrangian}
\end{equation}
where $\psi=(u,d)^T$,
$D_\mu=\partial_\mu-\ii\hat q A_\mu^{\rm ext}$,
$\hat q=\operatorname{diag}(2e/3,-e/3)$,
and $A_\mu^{\rm ext}=(0,0,Bx_1,0)$.
Here $m_0$ is the current light-quark mass,
$G_S$ is the coupling constant in the scalar and pseudoscalar channels,
and $\bm\tau$ denotes the Pauli matrices in flavor space.
We set $N_c=3$ and the quark-number chemical potential
to zero.

In the mean-field approximation, the constituent quark mass is defined as
\begin{equation}
M=m_0-2G_S\langle\bar\psi\psi\rangle .
\label{eq:constituent-mass}
\end{equation}
At finite temperature, magnetic field, and chiral chemical potential,
$M=M(T,B,\mu_5)$ is determined self-consistently from
\begin{equation}
\frac{\partial\Omega_{\rm MF}}{\partial M}=0 .
\label{eq:gap-stationary}
\end{equation}
Similar mean-field treatments of chirally imbalanced NJL matter can be
found in Refs.~\cite{Ghosh2024,Chaudhuri2022}.

For flavor $f=u,d$, we define
\begin{equation}
  B_f=|q_fB|,
  \qquad
  \varepsilon_{f,n}(p_3)
  =
  \sqrt{p_3^2+2nB_f}.
  \label{eq:transverse-energy}
\end{equation}

For the numerical calculation, it is convenient to fold the
longitudinal momentum integral onto $p_3\geq0$. We define
\begin{equation}
  E_{f,n,\eta}(p_3)
  =
  \sqrt{
    M^2+
    [\varepsilon_{f,n}(p_3)+\eta\mu_5]^2
  },
  \qquad
  \eta=\pm1,
  \label{eq:quark-branches}
\end{equation}
where, for $n\geq1$, $\eta=\pm1$ labels the two
quasiparticle branches. For the lowest Landau level (LLL), the two terms instead represent
the positive- and negative-$p_3$ sectors of the single independent
spin state after folding the full momentum axis.

The gap equation can then be written as
\begin{equation}
\begin{aligned}
M ={}& m_0
+2G_S N_c M
\sum_{f=u,d}\frac{B_f}{2\pi}
\sum_{n=0}^{\infty}
(2-\delta_{n0})
\sum_{\eta=\pm1}
\\
&\times
\int_0^\infty
\frac{\dd p_3}{2\pi}
\frac{
1-2n_F(E_{f,n,\eta})
}{
E_{f,n,\eta}
}.
\end{aligned}
\label{eq:gap-equation}
\end{equation}
Here $n_F(E)=[\exp(E/T)+1]^{-1}$. For each $(T,B,\mu_5)$, Eq.~\eqref{eq:gap-equation} is solved
self-consistently to obtain $M(T,B,\mu_5)$.

The Ritus propagator below is kept in its full-axis form, and the
half-axis representation is introduced only after folding the
longitudinal integrands.

In the presence of a uniform magnetic field, the coordinate-space
quark propagator has the Ritus representation
\cite{Schwinger1951,Ritus1972}.
Following the Ritus construction of Ref.~\cite{Mei2026}, the quark
propagator for flavor $f$ can be written as
\begin{align}
  S_f(x,y)
  ={}&
  \sum_{n=0}^{\infty}
  \int\frac{\dd p_0\,\dd p_2\,\dd p_3}{(2\pi)^3}
  e^{-\ii\widetilde p\cdot(x-y)}
  \nonumber\\
  &\times
  \mathcal P_n^{(f)}(x_1,p_2)
  S_f(\bar p_n)
  \overline{\mathcal P}_n^{(f)}(y_1,p_2),
  \label{eq:ritus-propagator}
\end{align}
where
\begin{align}
  \widetilde p^\mu
  &=
  (p_0,0,p_2,p_3),
  \nonumber\\
  \bar p_n^\mu
  &=
  \left(
    p_0,0,-s_f\sqrt{2nB_f},p_3
  \right),
  \nonumber\\
  s_f
  &=
  \sgn(q_fB).
  \label{eq:ritus-momenta}
\end{align}
The Ritus matrix is defined as
\begin{equation}
  \mathcal P_n^{(f)}(x_1,p_2)
  =
  g_n^{s_f}(x_1,p_2)\Delta_f^+
  +
  I_n g_{n-1}^{s_f}(x_1,p_2)\Delta_f^- ,
  \label{eq:ritus-projector}
\end{equation}
with
\begin{align}
  \Delta_f^\pm
  &=
  \frac{
    1\pm\ii s_f\gamma^1\gamma^2
  }{2},
  \qquad
  I_n=1-\delta_{n0},
  \nonumber\\
  g_n^{s_f}(x_1,p_2)
  &=
  \phi_n\left(
    x_1-\frac{s_f p_2}{|q_fB|}
  \right),
  \label{eq:ritus-spin-functions}
\end{align}
and
\begin{equation}
  \phi_n(\zeta)
  =
  \frac{|q_fB|^{1/4}}
       {\sqrt{2^n n!\sqrt{\pi}}}
  e^{-|q_fB|\zeta^2/2}
  H_n\left(\sqrt{|q_fB|}\,\zeta\right).
  \label{eq:landau-wavefunction}
\end{equation}
Here $H_n$ denotes the Hermite polynomial, and
\begin{equation}
  \overline{\mathcal P}_n^{(f)}
  =
  \gamma^0
  \left(\mathcal P_n^{(f)}\right)^\dagger
  \gamma^0 .
  \label{eq:ritus-conjugate}
\end{equation}
In the presence of the chiral chemical potential, the momentum-space
quark propagator is defined by
\begin{equation}
  S_f(\bar p_n)
  =
  \left[
    \slashed{\bar p}_n
    -M
    +\mu_5\gamma^0\gamma^5
  \right]^{-1}.
  \label{eq:momentum-quark-propagator}
\end{equation}

The matrices $\mathcal P_n^{(f)}$ contain the Landau wave functions
and spin projectors. After folding the longitudinal momentum
integral, the pole energies are represented by
Eq.~\eqref{eq:quark-branches}. 

Ultraviolet divergences are regularized using a four-term
Pauli--Villars scheme~\cite{PauliVillars1949}. For a given quasiparticle branch, we define
\begin{equation}
  E_{f,n,\eta}^{(j)}
  =
  \sqrt{
    M^2+a_j\Lambda^2+
    [\varepsilon_{f,n}+\eta\mu_5]^2
  },
  \qquad j=0,1,2,3,
\end{equation}
with
\begin{equation}
  \begin{aligned}
    (a_0,a_1,a_2,a_3)&=(0,1,2,3),\\
    (c_0,c_1,c_2,c_3)&=(1,-3,3,-1).
  \end{aligned}
\end{equation}
Divergent loop quantities are regularized by the replacement
\begin{equation}
  \mathcal I
  \longrightarrow
  \mathcal I_{\rm PV}
  =
  \sum_{j=0}^{3}c_j\,\mathcal I
  \bigl(E_{f,n,\eta}^{(j)}\bigr).
\end{equation}
The same prescription is applied consistently to the gap equation
and to the mesonic polarization functions.

The model parameters are fixed using vacuum observables. Using
$\langle\bar u u\rangle=\langle\bar d d\rangle
\simeq(-0.25~{\rm GeV})^3$
and $f_\pi=0.093~{\rm GeV}$ in vacuum, we take
$m_0=0.005~{\rm GeV}$,
$G_S=3.44~{\rm GeV}^{-2}$, and
$\Lambda=1.127~{\rm GeV}$.

\begin{figure}[t]
  \centering
  \includegraphics[width=0.95\columnwidth]{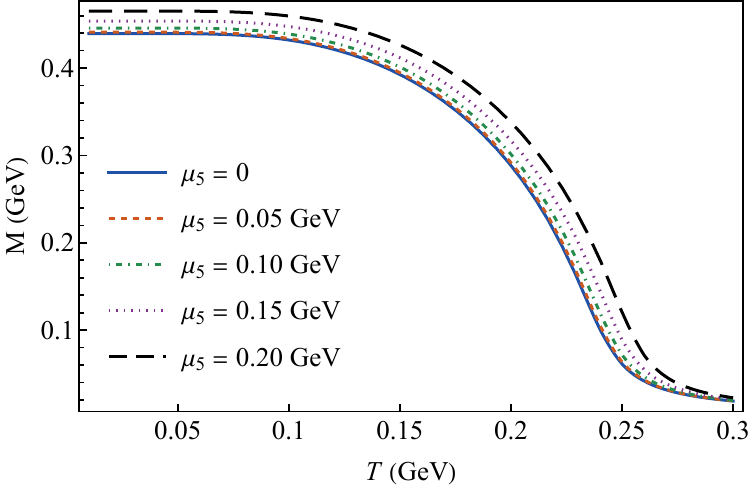}
  \caption{
  Constituent quark mass $M$ as a function of temperature $T$
  at fixed $eB=1~\mathrm{GeV}^2$ for several values of $\mu_5$.
  The mass is obtained self-consistently from
Eq.~\eqref{eq:gap-equation} using the parameters specified above.
  }
  \label{fig:quark-mass}
\end{figure}

As shown in Fig.~\ref{fig:quark-mass}, the constituent quark mass
decreases rapidly across the chiral crossover. Increasing $\mu_5$
slightly enhances $M$ at low temperature and shifts the crossover
region toward higher temperatures.
An increase of the transition temperature with $\mu_5$ has also been
reported in lattice simulations~\cite{Braguta2016Add} and in a
nonlocal NJL calculation~\cite{RuggieriPeng2016Add}.
These results provide a qualitative reference, although their
physical settings differ from the present magnetized calculation.
The sensitivity of the thermal transition to the regularization
prescription has been discussed in Ref.~\cite{Yu2016}.
The resulting self-consistent $M(T,B,\mu_5)$ is used throughout
the subsequent mesonic calculations.

\section{Neutral pion channel}
\label{sec:pi0}

\subsection{Polarization function}

We work in the single-channel RPA and neglect
$\sigma$--$\pi^0$ mixing.
The retarded neutral-pion propagator is
\begin{equation}
  D_{\pi^0}^R(\omega,\bm q)
  =
  \frac{2G_S}{1-2G_S\Pi_{\pi^0}^R(\omega,\bm q)}.
  \label{eq:pi0-rpa}
\end{equation}
The coordinate-space one-loop polarization is
\begin{equation}
  \Pi_{\pi^0}(x,y)
  =
  \ii N_c\sum_{f=u,d}
  \Tr_D\left[
    \ii\gamma^5S_f(x,y)
    \ii\gamma^5S_f(y,x)
  \right].
  \label{eq:pi0-coordinate}
\end{equation}
Here, $\Tr_D$ denotes the trace over the four-dimensional
Dirac spinor indices only. The color factor $N_c$ and
the flavor sum are written explicitly outside the trace.

For the neutral pion, translational invariance of the
polarization function is preserved. We therefore work
in momentum space and restrict the analytic reduction
to the rest frame,
\begin{equation}
  q^\mu=(\omega,\bm0).
  \label{eq:pi0-rest-frame}
\end{equation}

Before the Pauli--Villars sum is applied, the physical contribution to the retarded polarization function can be
written in the folded representation as
\begin{align}
  \Ree\Pi_{\pi^0}^{R}(\omega)
  ={}&\frac{N_c}{4\pi^2}
  \sum_{f=u,d}B_f
  \sum_{n=0}^{\infty}(2-\delta_{n0})
  \sum_{\eta=\pm1}
  \mathcal P\!\int_0^\infty\dd p_3
  \nonumber\\
  &\times
  \frac{4E_{f,n,\eta}
  \tanh[E_{f,n,\eta}/(2T)]}
  {4E_{f,n,\eta}^{2}-\omega^2}.
  \label{eq:pi0-polarization}
\end{align}
The regularized polarization used in the numerical calculation is
obtained by applying the Pauli--Villars prescription defined in
Sec.~\ref{sec:model} to Eq.~\eqref{eq:pi0-polarization}.

With the same Pauli--Villars prescription applied to the gap
equation and the neutral polarization function, the zero-energy
limit satisfies
\begin{equation}
   1-2G_S\Pi_{\pi^0}^{R}(0)=\frac{m_0}{M}.
  \label{eq:neutral-gap-identity}
\end{equation}
This relation fixes the neutral-loop normalization and its
chiral-limit behavior, and provides a direct consistency check on
the Landau-level and branch counting used in the neutral loop.

We follow the lowest continuous positive-energy zero of the
real inverse propagator and denote the selected solution by
$m_{\pi^0}$,
\begin{equation}
  \mathcal F_{\pi^0}(m_{\pi^0})
  \equiv
  1-2G_S\Ree\Pi_{\pi^0}^{R}(m_{\pi^0})
  =0,
  \qquad
  m_{\pi^0}>0.
  \label{eq:pi0-pole}
\end{equation}
We identify $m_{\pi^0}$ with the neutral-pion pole
mass when the selected simple zero lies below the
lowest physical pair-production threshold, $2M$.
In this region, the zero corresponds to an
isolated pole of the retarded propagator.
When the zero lies above the continuum onset,
the same equation defines a real-axis energy
solution rather than a true pole mass, since the
polarization function generally develops a
nonzero imaginary part.

\subsection{Branch decomposition and physical thresholds}

According to Eq.~(20), at vanishing external spatial
momentum, $\bm q=0$, the orthogonality of the transverse
Ritus eigenfunctions requires the two internal quark
propagators to have the same Landau-level index,
$n'=n$.
The explicit Dirac-trace calculation and transverse
momentum integration leading to this selection rule
are presented in Appendix~\ref{app:neutral-trace}.
Since the neutral-pion vertex is flavor diagonal and
the external longitudinal momentum vanishes, the two
internal quark lines also carry the same flavor and
longitudinal momentum $p_3$.
Consequently, they have the same effective Ritus
momentum $\varepsilon_{f,n}$.

To illustrate the branch structure, we adopt the
positive-half-axis representation consistent with
Eq.~\eqref{eq:pi0-polarization}, namely, $p_3\geq0$.
For $n\geq1$, we define the two quark energy branches as
\begin{align}
  \mathscr Q_{f,n,\eta}
  &=
  \frac{1-\eta\mathscr J_{f,n}}{2},
  \qquad \eta=\pm1,
  \nonumber\\
  \mathscr J_{f,n}
  &=
  \frac{\bm\Sigma\cdot\bm p_{f,n}}
       {\varepsilon_{f,n}},
  \label{eq:branch-projector}
\end{align}
where
\begin{equation}
  \bm p_{f,n}
  =
  \left(
    0,-s_f\sqrt{2nB_f},p_3
  \right),
  \qquad
  \Sigma^i=\gamma^5\gamma^0\gamma^i .
\end{equation}
Since $\mathscr J_{f,n}^2=1$, the projectors satisfy
\begin{equation}
  \mathscr Q_{f,n,\eta}
  \mathscr Q_{f,n,\eta'}
  =
  \delta_{\eta\eta'}
  \mathscr Q_{f,n,\eta},
  \qquad
  \sum_{\eta=\pm1}
  \mathscr Q_{f,n,\eta}=1 .
  \label{eq:projector-orthogonality}
\end{equation}

Using
\begin{equation}
  \bm\gamma\cdot\bm p_{f,n}
  =
  \varepsilon_{f,n}\gamma^0\gamma^5
  \mathscr J_{f,n},
\end{equation}
where
$\bm\gamma\cdot\bm p\equiv\sum_{i=1}^{3}\gamma^i p_i$,
together with
\begin{equation}
  \mathscr J_{f,n}\mathscr Q_{f,n,\eta}
  =
  -\eta\mathscr Q_{f,n,\eta},
\end{equation}
the inverse propagator in each branch becomes
\begin{equation}
  S_f^{-1}(\bar p_n)\mathscr Q_{f,n,\eta}
  =
  \left[
    \gamma^0p_0-M+
    (\eta\varepsilon_{f,n}+\mu_5)
    \gamma^0\gamma^5
  \right]
  \mathscr Q_{f,n,\eta}.
\end{equation}
Inverting this expression and using
$\sum_{\eta}\mathscr Q_{f,n,\eta}=1$ gives
\begin{equation}
  S_f(\bar p_n)
  =
  \sum_{\eta=\pm1}
  \mathscr Q_{f,n,\eta}
  \frac{
    \gamma^0p_0+M+
    (\eta\varepsilon_{f,n}+\mu_5)\gamma^0\gamma^5
  }{
    p_0^2-E_{f,n,\eta}^2
  } .
  \label{eq:branch-propagator-decomposition}
\end{equation}
Thus each Landau level with $n\geq1$ contains two quasiparticle
branches with poles at
\begin{equation}
  p_0=\pm E_{f,n,\eta} .
\end{equation}

To make the branch selection explicit, define
\begin{equation}
  \mathcal R_{f,n,\eta}(p_0)
  =
  \gamma^0p_0+M+
  (\eta\varepsilon_{f,n}+\mu_5)
  \gamma^0\gamma^5 .
  \label{eq:branch-numerator}
\end{equation}
The Dirac numerator associated with two internal branches
$\eta$ and $\eta'$ is then
\begin{align}
  \mathcal N_{f,n}^{\eta\eta'}
  (p_0,\omega)
  \equiv{}&
  \Tr_D\Big[
    \gamma^5
    \mathscr Q_{f,n,\eta}
    \mathcal R_{f,n,\eta}(p_0)
    \nonumber\\
    &\qquad\times
    \gamma^5
    \mathscr Q_{f,n,\eta'}
    \mathcal R_{f,n,\eta'}(p_0-\omega)
  \Big].
  \label{eq:branch-trace-definition}
\end{align}

Since
\begin{equation}
  [\gamma^5,\mathscr J_{f,n}]=0,
  \qquad
  [\gamma^0,\mathscr J_{f,n}]=0,
\end{equation}

the branch projectors commute with $\gamma^5$ and with the Dirac
numerators $\mathcal R_{f,n,\eta}$. They can therefore be brought
next to each other inside the trace:
\begin{align}
  \mathcal N_{f,n}^{\eta\eta'}
  ={}&
  \Tr_D\Big[
    \mathscr Q_{f,n,\eta}
    \mathscr Q_{f,n,\eta'}
    \gamma^5
    \mathcal R_{f,n,\eta}(p_0)
    \nonumber\\
    &\qquad\times
    \gamma^5
    \mathcal R_{f,n,\eta'}(p_0-\omega)
  \Big]
  \nonumber\\
  ={}&
  \delta_{\eta\eta'}
  \Tr_D\Big[
    \mathscr Q_{f,n,\eta}
    \gamma^5
    \mathcal R_{f,n,\eta}(p_0)
    \gamma^5
    \mathcal R_{f,n,\eta}(p_0-\omega)
  \Big].
  \label{eq:branch-selection}
\end{align}
Hence
\begin{equation}
  \mathcal N_{f,n}^{\eta\eta'}=0,
  \qquad
  \eta\neq\eta'.
  \label{eq:mixed-branch-vanishing}
\end{equation}
The explicit evaluation of the transverse-integrated
Dirac trace, including the cancellation of the
opposite-branch denominator factors, is presented
in Appendix~\ref{app:neutral-trace}.
The neutral pseudoscalar vertex therefore connects
only equal quasiparticle branches at zero external
spatial momentum.
The mixed-branch contribution vanishes in the complete
Dirac trace at fixed internal frequency and momentum,
before the Matsubara sum, longitudinal integration,
and Pauli--Villars subtraction.

The analytic consequence becomes transparent after factorizing
the two propagator denominators,
\begin{equation}
  p_0^2-E_{f,n,\eta}^2
  =
  (p_0-E_{f,n,\eta})
  (p_0+E_{f,n,\eta}).
\end{equation}
After the Matsubara sum and analytic continuation, a generic pair
of internal poles gives a denominator of the form
\begin{equation}
  \omega
  -sE_{f,n,\eta}
  +s'E_{f,n,\eta'},
  \qquad
  s,s'=\pm1 .
  \label{eq:generic-branch-denominator}
\end{equation}
Equation~\eqref{eq:mixed-branch-vanishing} eliminates all terms with
$\eta\neq\eta'$. Therefore, apparent mixed-branch denominators
involving $E_{f,n,+}$ and $E_{f,n,-}$ may occur at intermediate
stages of the calculation, but their complete Dirac residues vanish
and they do not generate physical poles or branch cuts.

\begin{figure*}[!t]
  \centering
  \begin{minipage}[t]{0.49\textwidth}
    \centering
    \includegraphics[width=\linewidth]{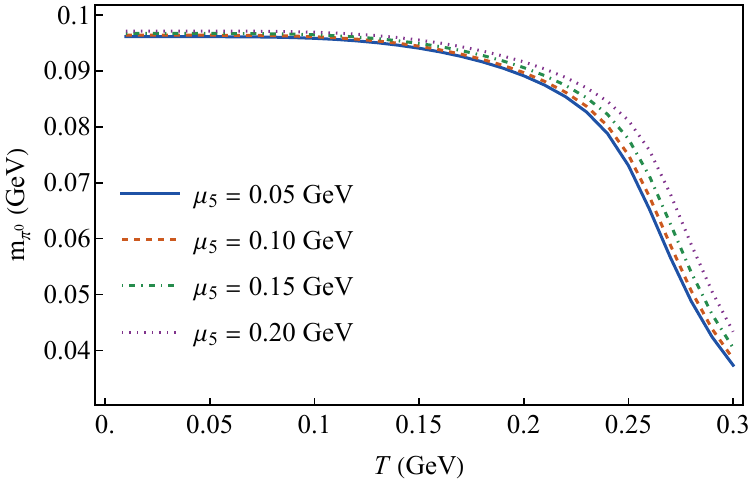}
    \vspace{-1.5ex}
    \par\small (a)
  \end{minipage}\hfill
  \begin{minipage}[t]{0.49\textwidth}
    \centering
    \includegraphics[width=\linewidth]{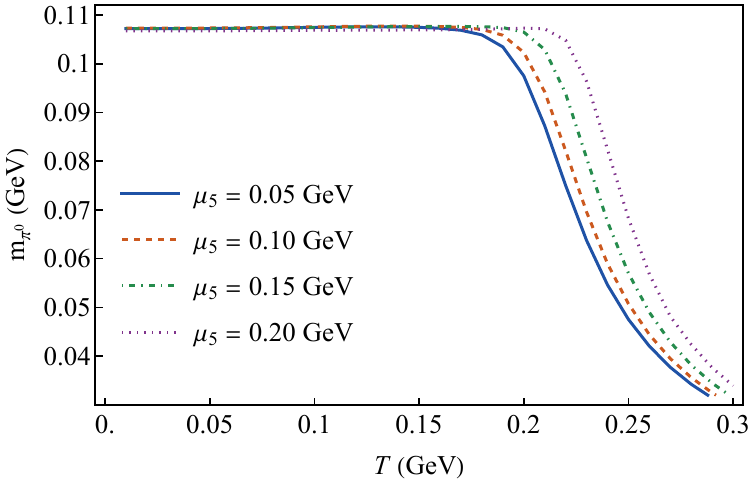}
    \vspace{-1.5ex}
    \par\small (b)
  \end{minipage}
  \caption{
  Lowest neutral-pion mass as a function of temperature
  for several values of $\mu_5$ at
  (a) $eB=1~\mathrm{GeV}^2$ and
  (b) $eB=0.4~\mathrm{GeV}^2$.
  At both magnetic fields, the $\mu_5$ dependence is weak in the
  low-temperature region, while the curves separate near the chiral
  crossover. Increasing $\mu_5$ shifts the rapid thermal decrease of
  the neutral-pion mass toward higher temperature.
  }
  \label{fig:pi0-temperature}
\end{figure*}

For quark--antiquark pair production, the two internal poles have
opposite energy signs, and the surviving singular denominators are
\begin{equation}
  \omega\pm2E_{f,n,\eta}.
  \label{eq:surviving-denominators}
\end{equation}
The same-branch scattering combination instead contains
$E_{f,n,\eta}-E_{f,n,\eta}=0$. At $\bm q=0$ its corresponding
thermal occupation difference also vanishes. Thus no finite-energy
Landau cut is generated in the neutral channel at zero external
spatial momentum. For the LLL, the two $\eta$ terms represent the folded
positive- and negative-$p_3$ sectors rather than independent
quasiparticle branches. At $q_3=0$, longitudinal momentum
conservation does not mix these two sectors, so the LLL does
not generate an additional finite-energy Landau cut either.

This branch selection is specific to the neutral pseudoscalar
channel at $\bm q=0$. It does not extend directly to
$\bm q\neq0$ or to the charged pion, for which the two internal
quark lines are not diagonal in the same Ritus basis.

Within the present mean-field and single-channel
RPA treatment, no mixed-branch finite-energy Landau
cut is generated at $B=0$ either.
This differs from the contribution reported in
Ref.~\cite{Ghosh2024}.
Here the conclusion follows directly from the
orthogonality of the common-momentum branch projectors
in the complete Dirac trace.

For $n\geq1$, the onset of the Unitary cut associated with a
given quasiparticle branch is the minimum energy required to create
a quark--antiquark pair,
\begin{equation}
  \omega_{f,n,\eta}^{\mathrm U}
  =
  2\min_{p_3\geq0}
  E_{f,n,\eta}(p_3),
  \qquad n\geq1.
  \label{eq:unitary-threshold-definition}
\end{equation}
Defining
\begin{equation}
  \lambda_{f,n}
  =
  \sqrt{2nB_f},
\end{equation}
one obtains
\begin{align}
  \omega_{f,n,+}^{\mathrm U}
  &=
  2\sqrt{
    M^2+
    (\lambda_{f,n}+\mu_5)^2
  },
  \label{eq:threshold-plus}
  \\
  \omega_{f,n,-}^{\mathrm U}
  &=
  2
  \begin{cases}
  M,
  & \mu_5\geq\lambda_{f,n},
  \\[2pt]
  \sqrt{
    M^2+
    (\lambda_{f,n}-\mu_5)^2
  },
  & \mu_5<\lambda_{f,n}.
  \end{cases}
  \label{eq:threshold-minus}
\end{align}

For the upper branch the minimum remains at $p_3=0$,
whereas the lower branch reaches $E=M$ at
$p_3=\sqrt{\mu_5^2-\lambda_{f,n}^2}$ when
$\mu_5\geq\lambda_{f,n}$. The corresponding threshold therefore
reduces to $2M$ in this regime.

On the real axis, singular momentum integrals contributing to
$\Ree\Pi_{\pi^0}^{R}$ are evaluated in the Cauchy principal-value
sense, whereas the imaginary part is determined by the corresponding
on-shell discontinuity.

The LLL requires no separate branch analysis. In the folded
representation, its lower-energy sector reaches $E=M$ at
$p_3=\mu_5$ for the positive $\mu_5$ considered here. Hence the
lowest physical neutral continuum onset remains
\begin{equation}
  \omega_{\rm th}^{\pi^0}=2M .
  \label{eq:lowest-pi0-threshold}
\end{equation}
There is no additional explicit $\mu_5$ shift in this
expression, but the threshold still depends on $\mu_5$
through the self-consistent mass $M(T,B,\mu_5)$.
Higher-Landau-level thresholds retain an explicit
branch-dependent $\mu_5$ dependence.

\subsection{Numerical results}
\label{subsec:pi0-results}

Earlier NJL studies have investigated the neutral-pion
pole mass and screening mass in magnetized matter
\cite{Avancini2017,Sheng2021}.
Here, we investigate the dependence of the neutral-pion
mass on temperature, magnetic field, and chiral imbalance. 

\begin{figure}[!t]
  \centering
  \includegraphics[width=\columnwidth]{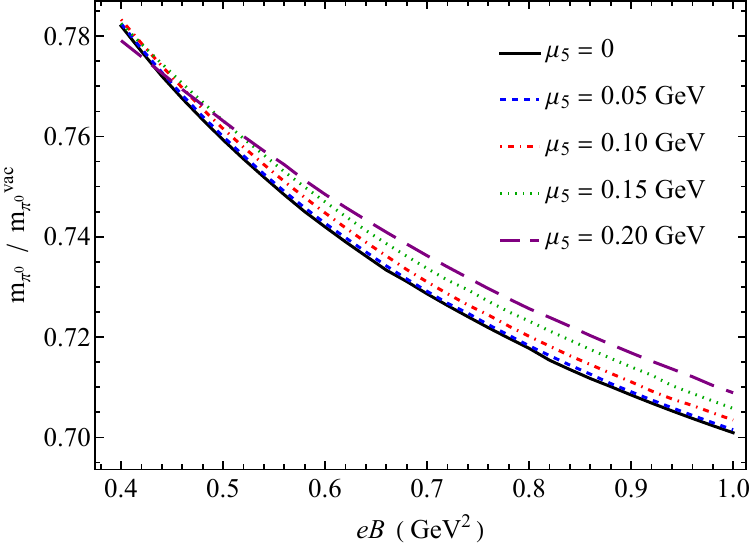}
  \caption{
  Ratio of the neutral-pion mass to its
  vacuum value,
  $m_{\pi^0}/m_{\pi^0}^{\rm vac}$, as a function of the magnetic field
  at $T=0$.
  Here, $m_{\pi^0}^{\rm vac}=0.134~\mathrm{GeV}$ denotes the vacuum pole mass at
  $T=B=\mu_5=0$.
  The curves correspond to the indicated values of $\mu_5$.
  }
  \label{fig:pi0-B-mu5}
\end{figure}

Figure~\ref{fig:pi0-temperature} shows the neutral-pion
mass as a function of temperature at
$eB=0.4~\mathrm{GeV}^2$ and $eB=1~\mathrm{GeV}^2$.
The $\mu_5$ dependence is weak in the low-temperature
region, while the curves separate more clearly near
the chiral crossover.
Increasing $\mu_5$ shifts the rapid decrease toward
higher temperatures, consistently with the shift
of the crossover in the self-consistent quark mass.

The $\mu_5$ dependence reflects both the modification
of the self-consistent chiral background and the
explicit $\mu_5$ dependence of the polarization function.

Figure~\ref{fig:pi0-B-mu5} shows the magnetic-field
dependence of the neutral-pion mass at $T=0$.
For every fixed $\mu_5$, the neutral-pion mass decreases
monotonically with increasing magnetic field over the
displayed interval.
The overall magnetic suppression is considerably
stronger than the comparatively small separation
among the different $\mu_5$ curves.
Unlike the charged pion, the neutral pion carries no
center-of-mass Landau energy.
Its magnetic response therefore originates from the
Landau-quantized quark loop together with the
self-consistent modification of the chiral background.

Lattice QCD calculations have also found that
the neutral-pion mass decreases with increasing
magnetic field.
This trend was observed in quenched Wilson-fermion
calculations \cite{Bali2018Add} and in a physical-mass
$(2+1)$-flavor lattice study \cite{Ding2026}.
Our result for the neutral-pion mass is qualitatively
consistent with these lattice QCD findings.

\section{Charged pion channel}
\label{sec:picharged}

\subsection{Mesonic Ritus representation}

We consider the $\pi^+$ channel. Its coordinate-space
one-loop polarization is
\begin{equation}
  \Pi_{\pi^+}(x,y)
  =
  \ii N_c
  \Tr_{D,f}\left[
    \ii\gamma^5\tau_- S(x,y)
    \ii\gamma^5\tau_+ S(y,x)
  \right],
  \label{eq:piplus-coordinate}
\end{equation}
where
$S=\operatorname{diag}(S_u,S_d)$ and
$\tau_\pm=(\tau_1\pm\ii\tau_2)/\sqrt{2}$.

For the charged pion, the two constituent quarks carry different
electric charges, so the transverse structure is not diagonal in an
ordinary plane-wave basis. The composite field must instead be
projected onto mesonic Ritus eigenfunctions,
\begin{align}
  \Pi_{\pi^\pm,\ell}(\bar q)
  ={}&
  \int\dd^4x\,\dd^4y\,
  \mathbb F_{\ell,k_2}^{(\pm)*}(x)
  \Pi_{\pi^\pm}(x,y)
  \mathbb F_{\ell,k_2}^{(\pm)}(y),
  \label{eq:charged-ritus-transform}
  \\
  \bar q_\ell^\mu
  ={}&
  \left(
    q_0,0,
    -s_{\pi^\pm}\sqrt{(2\ell+1)|eB|},
    q_3
  \right).
  \label{eq:charged-ritus-momentum}
\end{align}
The overall volume factor associated with the
plane-wave directions is understood to be divided out.

The mesonic Ritus eigenfunction is defined as
\begin{equation}
  \mathbb F_{\ell,k_2}^{(\pm)}(x)
  =
  e^{-\ii(q_0x_0-k_2x_2-q_3x_3)}
  f_{\ell,k_2}^{(\pm)}(x_1),
  \label{eq:meson-ritus-function}
\end{equation}
where
\begin{equation}
f_{\ell,k_2}^{(\pm)}(x_1)
=
\phi_\ell\left(
  x_1-\frac{s_{\pi^\pm}k_2}{|eB|}
\right).
  \label{eq:meson-ritus-transverse}
\end{equation}
Here $\phi_\ell$ has the same normalized harmonic-oscillator form
as Eq.~\eqref{eq:landau-wavefunction}, with $|q_fB|$ replaced by
$|eB|$ for the charged meson, and
\begin{equation}
  s_{\pi^\pm}
  =
  \sgn(Q_{\pi^\pm}B),
  \qquad
  Q_{\pi^\pm}=\pm e.
\end{equation}
The retarded RPA propagator in the $\ell$th mesonic Landau channel is
\begin{equation}
  D_{\pi^\pm,\ell}^{R}(\bar q)
  =\frac{2G_S}{1-2G_S\Pi_{\pi^\pm,\ell}^{R}(\bar q)}.
  \label{eq:charged-rpa}
\end{equation}
We focus on the $\pi^+$ channel in the lowest mesonic
Ritus level, $\ell=0$, at zero longitudinal momentum,
$q_3=0$.
The charged-pion mass is defined directly from the
corresponding lowest-channel energy,
$m_{\pi^+}\equiv\omega_{\pi^+,0}$.

\subsection{polarization function and analytic structure }

For the positive-half-axis representation, $p_3\geq0$, we define
\begin{equation}
  p_u
  =
  \sqrt{p_3^2+2n'B_u},
  \qquad
  p_d
  =
  \sqrt{p_3^2+2nB_d}.
  \label{eq:charged-longitudinal-momenta}
\end{equation}
The corresponding quasiparticle energies are
\begin{align}
  E_{u,n',\eta}
  &=
  \sqrt{M^2+(p_u+\eta\mu_5)^2},
  \nonumber\\
  E_{d,n,\eta}
  &=
  \sqrt{M^2+(p_d+\eta\mu_5)^2},
  \qquad
  \eta=\pm1.
  \label{eq:charged-quark-branches}
\end{align}

For the $u$-quark ($d$-quark) line, $\eta=\pm1$ labels the two
quasiparticle branches for $n'\geq1$ ($n\geq1$).

Because $|q_u|\neq|q_d|$, the $u$- and $d$-quark Ritus wave
functions involve different transverse magnetic scales. After
projection onto the charged-meson Ritus mode, their transverse
overlap is therefore generally non-diagonal in the quark Landau
indices $(n,n')$, and the charged polarization contains an
independent double Landau-level sum.

After the Matsubara sum, analytic continuation
$\ii\nu_m\to\omega+\ii0^+$, and folding the longitudinal momentum
integral onto $p_3\geq0$, the polarization function in the lowest
mesonic Ritus channel can be written as
\begin{align}
  \Pi_{\pi^+,0}^{R}(\omega)
  ={}&
  \mathcal N_B
  \sum_{n,n'=0}^{\infty}
  \int_0^\infty\dd p_3
  \nonumber\\
  &\times
  \left[
    j_{nn'}^{++}\mathcal K_{nn'}^{(1)}
    -
    j_{nn'}^{--}\mathcal K_{nn'}^{(2)}
  \right].
  \label{eq:charged-two-overlaps}
\end{align}
Here,
\begin{equation}
  \mathcal N_B
  =
  \frac{|eB|}{\pi^2}
  \left(\frac{2}{\pi}\right)^{3/2}.
  \label{eq:charged-overlap-normalization}
\end{equation}

The transverse coefficients $j_{nn'}^{++}$ and $j_{nn'}^{--}$
are defined in the normalization adopted here.
Their transverse wave-function structure follows
Ref.~\cite{Mei2026}, but their definition does not include
the factor $8N_c$ appearing in Eq.~(18) of that reference.
The overall prefactor $\mathcal N_B$ in
Eq.~\eqref{eq:charged-overlap-normalization}
already includes the color multiplicity $N_c=3$.

The longitudinal kernels
$\mathcal K_{nn'}^{(1)}$ and
$\mathcal K_{nn'}^{(2)}$
contain the quasiparticle energies, thermal occupation factors,
and pole residues of both $\eta$ sectors and of the positive-
and negative-energy quark poles. Their explicit pole-residue
forms are collected in Appendix~\ref{app:charged-reduction}.

\begin{figure}[t]
  \centering
  \begin{tikzpicture}[
    x=0.72cm,
    y=0.85cm,
    >=Latex,
    font=\footnotesize
  ]

  \begin{scope}[yshift=1.6cm]

    \draw[->] (-4.6,0) -- (4.7,0)
      node[right] {$\Re\,\omega$};

    \draw[->] (0,-0.35) -- (0,0.85)
      node[above] {$\Im\,\omega$};

    \draw[very thick,<-]
      (-2.6,0) -- (-4.2,0);

    \draw[very thick,->]
      (2.6,0) -- (4.2,0);

    \draw (-2.6,-0.10) -- (-2.6,0.10);
    \draw ( 2.6,-0.10) -- ( 2.6,0.10);

    \node[below] at (-2.6,-0.10)
      {$-\omega_{nn'}^{\rm U}$};

    \node[below] at (2.6,-0.10)
      {$\omega_{nn'}^{\rm U}$};

    \node[above] at (-3.4,0.08)
      {Unitary};

    \node[above] at (3.4,0.08)
      {Unitary};

  \end{scope}

  \begin{scope}[yshift=-0.35cm]

    \draw[->] (-4.6,0) -- (4.7,0)
      node[right] {$\Re\,\omega$};

    \draw[->] (0,-0.35) -- (0,0.85)
      node[above] {$\Im\,\omega$};

    \draw[very thick]
      (-2.0,0) -- (2.0,0);

    \draw (-2.0,-0.10) -- (-2.0,0.10);
    \draw ( 2.0,-0.10) -- ( 2.0,0.10);

    \node[below] at (-2.0,-0.10)
      {$-\omega_{nn'}^{\rm L}$};

    \node[below] at (2.0,-0.10)
      {$\omega_{nn'}^{\rm L}$};

    \node[above] at (0.85,0.08)
      {Landau};

  \end{scope}

  \end{tikzpicture}

  \caption{
  Schematic analytic structure of the charged-pion polarization
  for a fixed Landau-level pair $(n,n')$ at $\mu_5=0$.
  The Unitary cuts start at
  $\omega_{nn'}^{\rm U}=m_{u,n'}+m_{d,n}$,
  while the Landau cut extends between
  $\pm\omega_{nn'}^{\rm L}$, with
  $\omega_{nn'}^{\rm L}=|m_{u,n'}-m_{d,n}|$.
  }
  \label{fig:charged-cut-schematic}
\end{figure}
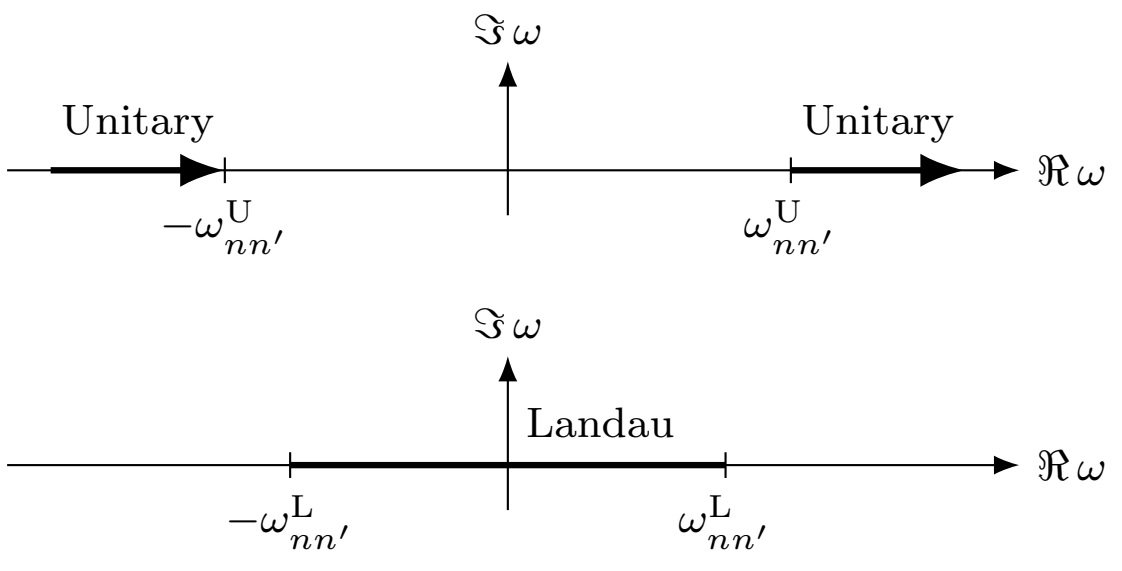

The analytic structure of the charged-pion retarded polarization
function is governed by its on-shell quasiparticle denominators.
For positive external energy, the two relevant singularity
conditions are
\begin{align}
  \omega
  &=
  E_{u,n',\eta}
  +
  E_{d,n,\eta'},
  \label{eq:charged-unitary-process}
  \\
  \omega
  &=
  \left|
    E_{u,n',\eta}
    -
    E_{d,n,\eta'}
  \right|.
  \label{eq:charged-landau-process}
\end{align}
The first condition describes quark--antiquark pair
creation, $\pi^+\leftrightarrow u+\bar d$.
The second describes thermal absorption or emission,
such as $\pi^++d\leftrightarrow u$ and
$\pi^++\bar u\leftrightarrow\bar d$.
These processes contribute to the Unitary and Landau
cuts, respectively, when their transverse Ritus
overlaps and complete on-shell residues are nonzero.

For each contributing Landau-level and branch sector,
the minimum of
$E_{u,n',\eta}+E_{d,n,\eta'}$
at a common longitudinal momentum determines the
pair-production onset.
The range of
$|E_{u,n',\eta}-E_{d,n,\eta'}|$
determines the corresponding scattering continuum.
Different Landau-level pairs generally produce
different boundaries.

At $\mu_5=0$, this picture reduces to the familiar purely magnetic
result. The two $\eta$ sectors become degenerate, and
\begin{align}
  \omega_{nn'}^{\mathrm U}
  &=
  m_{u,n'}
  +
  m_{d,n},
  \nonumber\\
  \mathcal C_{nn'}^{\mathrm L}
  &=
  \left[
    0,
    |m_{u,n'}-m_{d,n}|
  \right],
  \label{eq:charged-zero-mu5-cuts}
\end{align}
where
\begin{equation}
  m_{u,n'}
  =
  \sqrt{M^2+2n'B_u},
  \qquad
  m_{d,n}
  =
  \sqrt{M^2+2nB_d}.
  \label{eq:charged-landau-masses}
\end{equation}
The corresponding analytic structure is illustrated schematically
in Fig.~\ref{fig:charged-cut-schematic}, in agreement with the
purely magnetic result of Ref.~\cite{Mei2026}.

\begin{figure*}[!t]
  \centering
  \begin{minipage}[t]{0.325\textwidth}
    \centering
    \includegraphics[width=\linewidth]{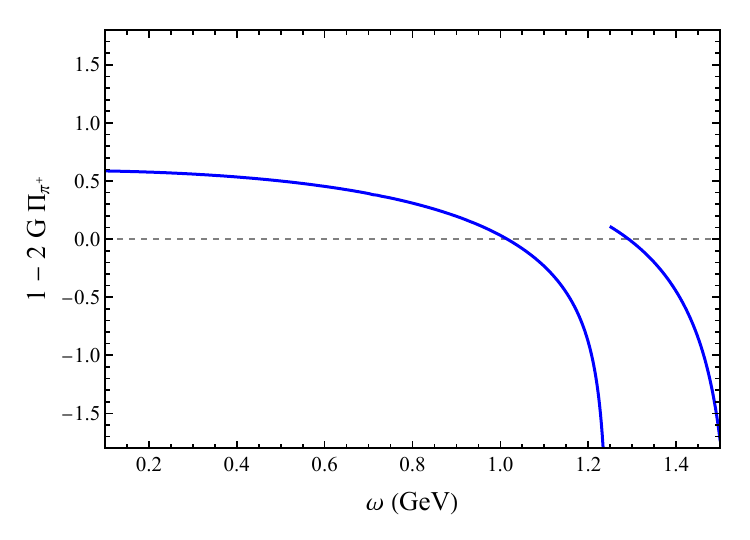}
    \vspace{-1.2ex}
    \par\small (a)
  \end{minipage}\hfill
  \begin{minipage}[t]{0.325\textwidth}
    \centering
    \includegraphics[width=\linewidth]{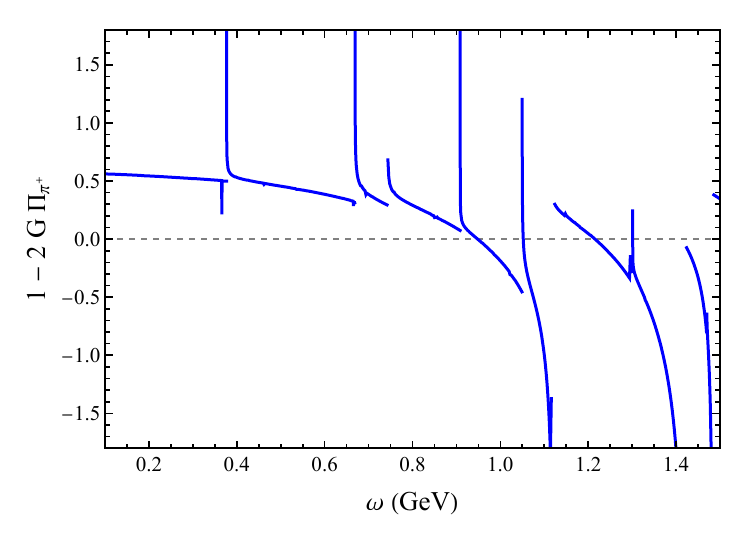}
    \vspace{-1.2ex}
    \par\small (b)
  \end{minipage}\hfill
  \begin{minipage}[t]{0.325\textwidth}
    \centering
    \includegraphics[width=\linewidth]{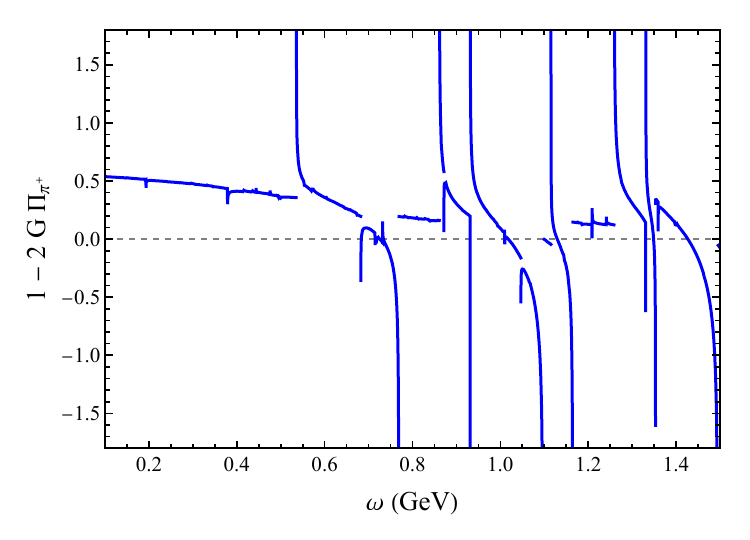}
    \vspace{-1.2ex}
    \par\small (c)
  \end{minipage}
  \caption{
  Real inverse propagator
  $\mathcal F_{\pi^+}(\omega)$
  in the lowest mesonic Ritus channel at
  $eB=1~\mathrm{GeV}^2$,
  $\mu_5=0.20~\mathrm{GeV}$,
  and $q_3=0$.
  Panels (a)--(c) correspond to
  $T=0$, $0.18$, and $0.25~\mathrm{GeV}$,
  respectively.
  Blue curves show the Pauli--Villars-regularized
  principal-value result, and the gray dashed line denotes
  $\mathcal F_{\pi^+}=0$.
  The quasiparticle branch is identified by the continuous
  downward crossing specified in
  Eq.~\eqref{eq:piplus-physical-mass-criterion}.
  }
  \label{fig:piplus-inverse-scans}
\end{figure*}

\begin{figure}[!t]
  \centering
  \includegraphics[width=\columnwidth]{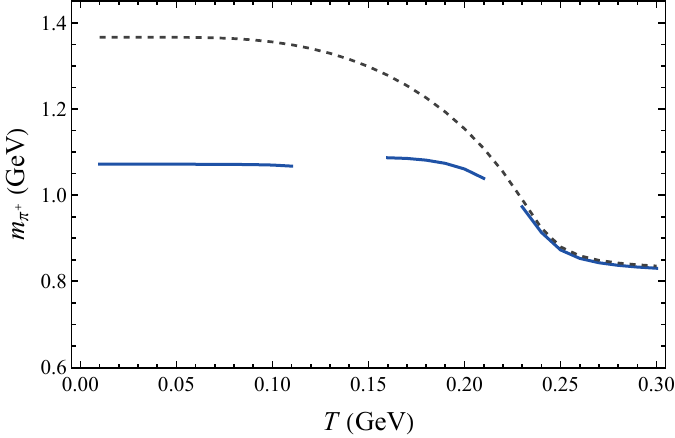}
  \caption{Charged-pion mass in the lowest mesonic Ritus channel as a function of temperature at $\mu_5=0$ and $eB=1~\mathrm{GeV}^2$ (blue segments), compared with
  the lowest physical quark--antiquark Unitary threshold
  in the same truncation (gray dashed line).
  Breaks separate the displayed pion-mass solution segments.
  }

  \label{fig:piplus-mu5-0}
\end{figure}

For the higher-Landau-level sectors, a nonzero $\mu_5$ splits
the quasiparticle branches. The upper
branch remains monotonic in $p_3$, whereas a lower branch may
develop its minimum at nonzero longitudinal momentum when
$\mu_5$ exceeds the corresponding Landau scale. Since the
$u$- and $d$-quark Landau scales differ, their minima need not
occur at the same $p_3$. The lowest charged pair-production
threshold must therefore be determined from the full
$u$--$d$ dispersion and has no universal form analogous to the
neutral value $2M$.

At large $p_3$, the absolute energy differences
between equal-$\eta$ sectors approach zero, whereas
those between opposite-$\eta$ sectors approach
$2\mu_5$.
The latter is an asymptotic splitting scale rather
than a universal Landau-cut endpoint.

\subsection{charged-pion results}
\label{subsec:charged-poles}

For the lowest charged-pion Ritus channel, we define
the real inverse propagator
\begin{equation}
  \mathcal F_{\pi^+}(\omega)
  =
  1-2G_S\Ree\Pi_{\pi^+,0}^{R}(\omega).
\end{equation}

The charged-pion mass is defined by the selected
positive-energy solution of
\begin{equation}
  \mathcal F_{\pi^+}(m_{\pi^+})=0,
  \qquad
  m_{\pi^+}>0,
  \label{eq:piplus-mass-equation}
\end{equation}
where $m_{\pi^+}\equiv\omega_{\pi^+,0}$ denotes the
pion mass in the lowest mesonic Ritus channel at
$q_3=0$.

Starting from the zero-temperature pion solution,
we follow the selected root as the temperature varies.
At each temperature, an admissible solution must lie
within a continuous interval of
$\mathcal F_{\pi^+}(\omega)$ and satisfy the local
slope criterion below.
Accordingly, a sign change across a cut endpoint
or a divergent boundary is not identified with a
physical solution of Eq.~\eqref{eq:piplus-mass-equation}.

With the propagator convention adopted here, we define
\begin{equation}
  \mathcal N_{\pi^+}
  =
  -\left.
  \frac{\partial\mathcal F_{\pi^+}}
       {\partial\omega^2}
  \right|_{\omega=m_{\pi^+}} .
  \label{eq:piplus-norm}
\end{equation}
A positive $\mathcal N_{\pi^+}$ specifies a downward
crossing, with the same local slope sign as the
zero-temperature pion solution.
The selected pion mass therefore satisfies
\begin{equation}
  \begin{aligned}
  &\mathcal F_{\pi^+}(m_{\pi^+})=0,
  \\
  &\mathcal N_{\pi^+}>0,
  \\
  &\mathcal F_{\pi^+}
  \ \text{is continuous near}\ m_{\pi^+}.
  \end{aligned}
  \label{eq:piplus-physical-mass-criterion}
\end{equation}

\begin{figure*}[!t]
  \centering
  \includegraphics[width=0.88\textwidth]{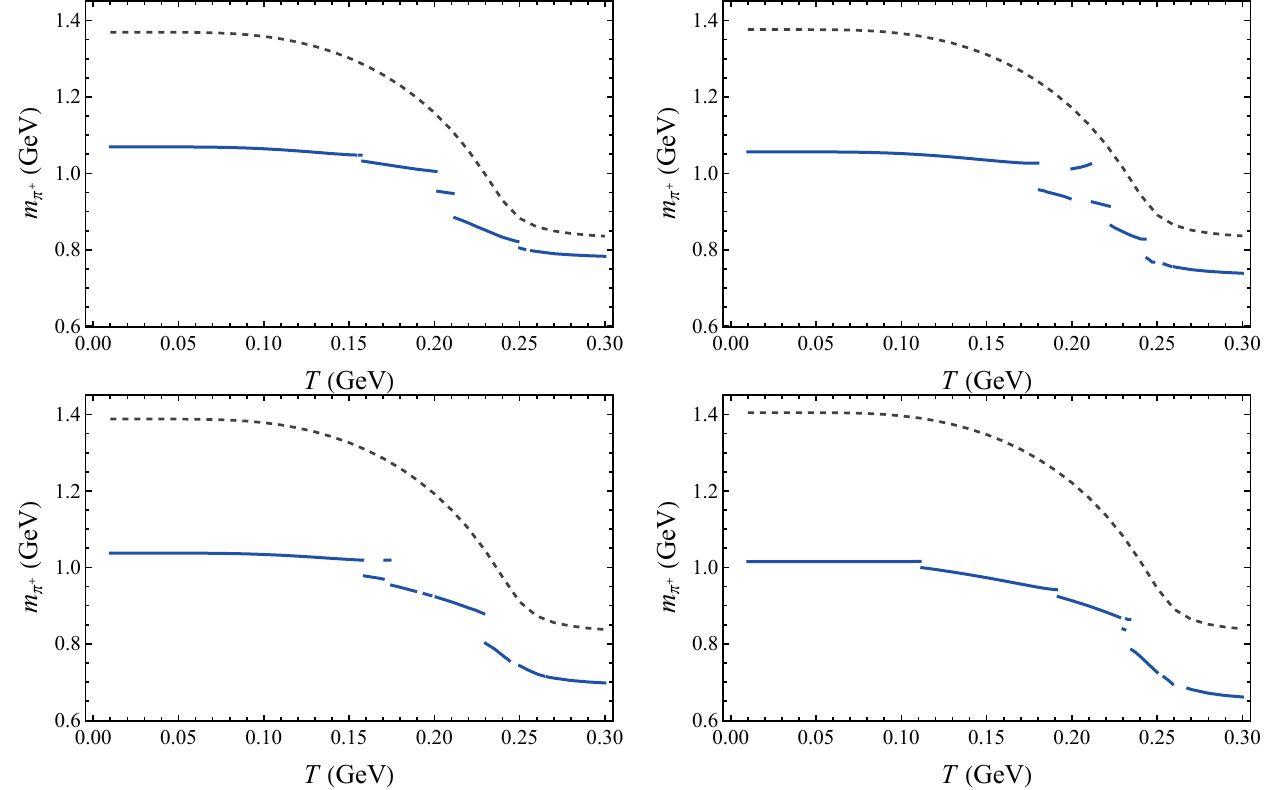}
  \caption{Charged-pion mass (blue segments) and the lowest
  physical quark--antiquark Unitary threshold (gray dashed line) in the lowest mesonic Ritus channel at
  $eB=1~\mathrm{GeV}^2$.
  From the upper-left to the lower-right panel,
  $\mu_5=0.05$, $0.10$, $0.15$, and $0.20~\mathrm{GeV}$.
  Each mass solution satisfies
  Eq.~\eqref{eq:piplus-physical-mass-criterion}.
  }

  \label{fig:piplus-threshold}
\end{figure*}

Being below the lowest Unitary threshold excludes the
corresponding quark--antiquark production channel,
but does not by itself guarantee stability because
thermal Landau absorption may still contribute.
At low temperature, these thermal contributions are
Fermi suppressed.

At $T=0$, Fig.~\ref{fig:piplus-inverse-scans}(a)
exhibits a relatively simple real-axis structure because
the thermal Landau contributions vanish.
The lowest zero crossing continuously tracked along
the real axis defines the meson mass in this limit
and provides a reference for determining the meson
mass at finite temperature.

The higher-temperature scans in
Figs.~\ref{fig:piplus-inverse-scans}(b) and
\ref{fig:piplus-inverse-scans}(c) show more structure near the
pair-production and thermal-scattering boundaries. The reduction
of $M$ changes the pair-production energies, while thermal
occupation makes scattering processes among the allowed
$u$--$d$ Landau sectors increasingly relevant. These processes
reshape the real part of the correlator and can separate its zeros
into distinct intervals. 

The dashed curves in Figs.~\ref{fig:piplus-mu5-0} and
\ref{fig:piplus-threshold} denote the lowest pair-production threshold
among the Landau-level and $\eta$ sectors with nonvanishing Ritus
overlap and on-shell residue.

In Fig.~\ref{fig:piplus-mu5-0}, the lowest pair-production
threshold generally shifts downward as the constituent quark mass decreases
with temperature. Above this threshold, the decay
$\pi^+\to u+\bar d$ becomes kinematically allowed, and the
charged pion is unstable against quark--antiquark decay.
We therefore discard mass solutions above the lowest
pair-production threshold. This exclusion produces gaps in
the temperature dependence of the pion mass, leaving separate
segments within which the solutions are followed continuously.
The high-temperature branch approaches the pair-production
threshold near the chiral crossover.
Mass jumps associated with Mott transitions have been
reported in magnetized NJL calculations~\cite{Mao2019}.

In Fig.~\ref{fig:piplus-threshold}, increasing $\mu_5$
lowers the mass over the intervals shown. The
$\mu_5$-dependent $u$- and $d$-quark dispersions also shift the
lowest allowed pair-production threshold, but this kinematic change
alone does not determine the pion mass because the chiral
background and the charged $u$--$d$ Ritus overlaps enter the full
correlator. The selected solution remains below the displayed
Unitary threshold over the plotted segments, so the corresponding
pair-production channel is kinematically closed, although thermal
scattering may still generate a width. The temperature intervals spanned by the selected real-axis
solution segments also vary with $\mu_5$.

\begin{figure}[t]
  \centering
  \includegraphics[width=\columnwidth]{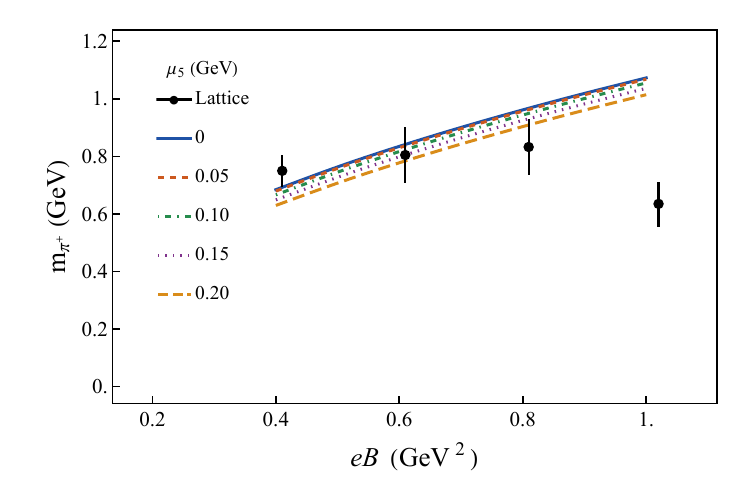}
 \caption{Charged-pion mass $m_{\pi^+}$ in the lowest mesonic
  Ritus channel as a function of $eB$ at $T=0$.
  The colored curves show the NJL results for the
  indicated values of $\mu_5$.
  The black points show selected continuum-extrapolated
  zero-temperature lattice QCD results from
  Ref.~\cite{Ding2026}, obtained without an external
  axial source.
  The $\mu_5=0$ model curve provides the baseline
  for comparison, while the finite-$\mu_5$ curves
  show the response to an imposed source.
  }

  \label{fig:piplus-B-scan}
\end{figure}

Figure~\ref{fig:piplus-B-scan} shows the calculated magnetic-field
dependence of the charged-pion mass.
Previous studies have investigated charged-meson spectroscopy and
structure in strong magnetic fields, while
zero-temperature lattice QCD exhibits a nonmonotonic magnetic
dependence of the charged-pion ground-state energy
\cite{Ding2021}.
We therefore use the $\mu_5=0$
model curve as the baseline for comparison, while the
finite-$\mu_5$ curves describe the response to an imposed
axial source.
Both sets of results
refer to $T=0$.

For comparison, a pointlike charged scalar in the
lowest Landau level would satisfy
\begin{equation}
  m_{\pi^+}^2(B)
  =
  m_{\pi^+}^2(0)+|eB|.
\end{equation}
Deviations from this behavior provide a qualitative
measure of interaction and internal-structure effects.
Chiral perturbation theory
and lattice polarizability calculations
~\cite{Luschevskaya2016Add} provide examples of
additional pion-structure contributions beyond
the pointlike Landau behavior.

Every fixed-$\mu_5$ curve in
Fig.~\ref{fig:piplus-B-scan} increases throughout
the displayed field interval, whereas increasing
$\mu_5$ at fixed magnetic field lowers the
charged-pion mass.
Zero-temperature lattice studies with different
lattice spacings and quark masses instead find
a weak-field increase followed by saturation or
a decrease at intermediate-to-strong fields.

The decrease of the charged-pion mass with increasing
$\mu_5$ suggests a possible counteracting contribution
if the effective chiral imbalance increases with
the magnetic field.
Whether this contribution can produce a downturn
depends on the relation $\mu_5(B)$
\cite{Zhai2026Review}.
Testing this possibility at $T=0$ requires a specified
relation $\mu_5(B)$, which is not determined by
the present fixed-source calculation.

\section{Conclusions}
\label{sec:conclusions}

We investigated neutral and charged pion correlations
in magnetized quark matter with a prescribed axial
source $\mu_5$.
At zero external spatial momentum, the complete
Dirac trace in the neutral channel eliminates
mixed-branch contributions, so that no finite-energy
Landau cut is generated.
The lowest neutral-pion pair-production threshold
remains $2M$, while the higher-Landau-level thresholds
split according to the quasiparticle branches.
In contrast, the charged channel involves
non-diagonal Landau-level couplings through the
Ritus overlaps and exhibits both pair-production
and thermal-scattering contributions.

The numerical results show that the neutral-pion mass
is weakly sensitive to $\mu_5$ at low temperature,
while its rapid thermal decrease shifts toward
higher temperatures as $\mu_5$ increases.
The charged-pion mass decreases with increasing
$\mu_5$ over the selected solution segments.
At zero temperature and fixed $\mu_5$, increasing
the magnetic field lowers the neutral-pion mass
but raises the charged-pion mass over the range studied.
These contrasting responses reflect the different
magnetic-field dependence of the neutral and charged
pion channels.

At $\mu_5=0$, the charged-pion mass increases
monotonically with the magnetic field in our calculation,
whereas lattice QCD reports a nonmonotonic field
dependence of the charged-pion ground-state energy.
The decrease of the charged-pion mass with increasing
$\mu_5$ found here provides an additional contribution
that may be relevant to this behavior.
Since $B$ and $\mu_5$ are treated as independent
external parameters, investigating their combined
effect along a specified relation $\mu_5(B)$
will be useful for further comparison with lattice results.

\begin{acknowledgments}
Mei Huang is supported in part by the National Natural Science Foundation of China (NSFC) Grant Nos: 12235016, 12221005 and Jie Mei is supported by the China Postdoctoral Science Foundation under Grant No. 2026M793760.
\end{acknowledgments}

\appendix

\section{Explicit derivation of the neutral-pion polarization function}
\label{app:neutral-trace}

In this appendix, we present the explicit algebraic
reduction of the neutral-pion polarization function
at zero external spatial momentum. The branch-selection
rule has been discussed in Sec.~\ref{sec:pi0} using
the Ritus projectors. Here, we provide the intermediate
Dirac-trace expressions and demonstrate explicitly
the cancellation of the opposite-branch denominator
factors. We use the same single-channel RPA
approximation as in the main text, with
$\sigma$--$\pi^0$ mixing neglected.

We denote the internal quark momentum by
$P=(p_0,\bm p)$ and the external pion momentum by
$Q=(K_0,\bm 0)$, where
\begin{equation}
p_0=i\omega_\ell,\qquad
K_0=i\nu_m,
\end{equation}
with
$\omega_\ell=(2\ell+1)\pi T$ and
$\nu_m=2m\pi T$.
The Dirac matrices are taken in the Weyl representation,
\begin{equation}
\gamma^0=
\begin{pmatrix}
0&I_2\\
I_2&0
\end{pmatrix},
\qquad
\gamma^i=
\begin{pmatrix}
0&\sigma_i\\
-\sigma_i&0
\end{pmatrix},
\qquad
\gamma^5=i\gamma^0\gamma^1\gamma^2\gamma^3,
\end{equation}
where $I_2$ is the two-dimensional identity matrix
and $\sigma_i$ are the Pauli matrices.
Throughout this appendix, $\Tr_D$ denotes the trace
over the four-dimensional Dirac indices.

Using
\begin{equation}
x_f=\frac{2p_\perp^2}{B_f},
\qquad
p_\perp^2=p_1^2+p_2^2,
\qquad
B_f=|q_fB|,
\end{equation}
the momentum-space Ritus propagator can be written as
\begin{equation}
S_f(P)=\sum_{n=0}^{\infty}S_{f,n}(P),
\end{equation}
where
\begin{equation}
S_{f,n}(P)
=
i(-1)^n e^{-x_f/2}
\frac{
\mathcal N_{f,n}^{\mathrm{prop}}(P)
\mathcal D_{f,n}(P)
}{
U_{f,n}(p_0)
}.
\label{eq:app-pi0-propagator}
\end{equation}
The numerator is
\begin{align}
\mathcal N_{f,n}^{\mathrm{prop}}(P)
={}&
2\mathcal A_f(P)
\left[
L_n(x_f)\Delta_f^+
-
L_{n-1}(x_f)\Delta_f^-
\right]
\nonumber\\
&+
4L_{n-1}^{1}(x_f)
(p_1\gamma^1+p_2\gamma^2),
\label{eq:app-pi0-N}
\end{align}
with
\begin{equation}
\mathcal A_f(P)
=
p_0\gamma^0-p_3\gamma^3+M
-\mu_5\gamma^0\gamma^5,
\end{equation}
and $L_{-1}=L_{-1}^{1}=0$.
Here, $\Delta_f^\pm$ are the spin projectors
defined in Eq.~\eqref{eq:ritus-spin-functions}.

The remaining matrix factor is
\begin{align}
\mathcal D_{f,n}(P)
={}&
\left(
p_0^2-\varepsilon_{f,n}^2-M^2+\mu_5^2
\right)I_4
\nonumber\\
&-2M\mu_5\gamma^0\gamma^5
+2p_0\mu_5\gamma^5,
\label{eq:app-pi0-D}
\end{align}
where $I_4$ is the four-dimensional identity matrix.
The scalar denominator reads
\begin{equation}
U_{f,n}(p_0)
=
\left(
p_0^2-\varepsilon_{f,n}^2-M^2-\mu_5^2
\right)^2
-4\mu_5^2\varepsilon_{f,n}^2.
\label{eq:app-pi0-U}
\end{equation}

At finite temperature, the continuous energy integral
is replaced by the fermionic Matsubara sum. With the
pseudoscalar-vertex and propagator conventions of
Eq.~\eqref{eq:pi0-coordinate}, the polarization function
becomes
\begin{align}
\Pi_{\pi^0}(K_0)
={}&
-N_c T\sum_{\ell}
\sum_{f=u,d}
\int\frac{\dd p_3}{2\pi}
\frac{\dd^2p_\perp}{(2\pi)^2}
\sum_{n_1,n_2=0}^{\infty}
\nonumber\\
&\times
\Tr_D\left[
\gamma^5 S_{f,n_1}(P+Q)
\gamma^5 S_{f,n_2}(P)
\right].
\label{eq:app-pi0-loop}
\end{align}

Substituting Eq.~\eqref{eq:app-pi0-propagator}
into Eq.~\eqref{eq:app-pi0-loop}, we define the
transverse-integrated numerator
\begin{align}
h_{f;n_1n_2}
={}&
\int\dd^2p_\perp\,
\Tr_D\Big[
\gamma^5
i(-1)^{n_1}e^{-x_f/2}
\nonumber\\
&\quad\times
\mathcal N_{f,n_1}^{\mathrm{prop}}(P+Q)
\mathcal D_{f,n_1}(P+Q)
\nonumber\\
&\quad\times
\gamma^5
i(-1)^{n_2}e^{-x_f/2}
\mathcal N_{f,n_2}^{\mathrm{prop}}(P)
\mathcal D_{f,n_2}(P)
\Big].
\label{eq:app-pi0-h-definition}
\end{align}
The polarization function can consequently be written as
\begin{align}
\Pi_{\pi^0}(K_0)
={}&
-\frac{N_cT}{(2\pi)^3}
\sum_{f=u,d}
\sum_{\ell=-\infty}^{\infty}
\int_{-\infty}^{\infty}\dd p_3
\nonumber\\
&\times
\sum_{n_1,n_2=0}^{\infty}
\frac{
h_{f;n_1n_2}
}{
U_{f,n_1}(p_0+K_0)
U_{f,n_2}(p_0)
}.
\label{eq:app-pi0-h-integrand}
\end{align}
The factor $N_c$ accounts for the color trace,
while the flavor sum and Matsubara sum are
kept explicitly.

The transverse momentum integration is performed
using the Laguerre-polynomial orthogonality relations
\begin{align}
\int_0^\infty e^{-x_f}
L_n(x_f)L_{n'}(x_f)\,\dd x_f
&=\delta_{nn'},\\
\int_0^\infty x_f e^{-x_f}
L_{n-1}^{1}(x_f)L_{n'-1}^{1}(x_f)\,\dd x_f
&=n\delta_{nn'},
 n,n'\geq1,
\end{align}
together with
\begin{equation}
\dd^2p_\perp
=
\frac{\pi B_f}{2}\dd x_f.
\end{equation}
After evaluating the Dirac trace, terms containing
a single transverse momentum component, $p_1$ or $p_2$,
vanish upon angular integration. The surviving
radial terms involve products of Laguerre polynomials
of the same order, whose orthogonality enforces
\begin{equation}
h_{f;n_1n_2}=0,
\qquad n_1\neq n_2.
\label{eq:app-pi0-landau-selection}
\end{equation}
Thus, the two internal quark lines have the same
Landau-level index, and the double Landau-level
sum reduces to a single sum.

For $n=n_1=n_2\geq1$, the result of the Dirac trace
and transverse momentum integration is
\begin{align}
h_{f;nn}
={}&
-4\pi B_f
\sum_{\eta=\pm1}
\left[
E_{f,n,\eta}^2
-p_0(p_0+K_0)
\right]
\nonumber\\
&\times
\left[
p_0^2-E_{f,n,-\eta}^2
\right]
\left[
(p_0+K_0)^2-E_{f,n,-\eta}^2
\right].
\label{eq:app-pi0-h-HLL}
\end{align}
Here, $E_{f,n,\eta}$ is the quasiparticle
dispersion defined in Eq.~\eqref{eq:quark-branches}.

The scalar denominator factorizes as
\begin{equation}
U_{f,n}(p_0)
=
\left(p_0^2-E_{f,n,+}^2\right)
\left(p_0^2-E_{f,n,-}^2\right).
\label{eq:app-pi0-U-factor}
\end{equation}
Combining the numerator with the two propagator
denominators gives
\begin{align}
&\frac{h_{f;nn}}
{U_{f,n}(p_0+K_0)U_{f,n}(p_0)}
\nonumber\\
={}&
-4\pi B_f
\sum_{\eta=\pm1}
\frac{
E_{f,n,\eta}^2-p_0(p_0+K_0)
}{
\left(p_0^2-E_{f,n,\eta}^2\right)
\left[(p_0+K_0)^2-E_{f,n,\eta}^2\right]
}.
\label{eq:app-pi0-same-branch}
\end{align}
The opposite-branch denominator factors cancel
between the numerator and denominator, leaving
only contributions with the same branch index
on the two quark propagators.

The lowest Landau level (LLL) contains only one
independent spin state. At fixed signed longitudinal
momentum, its two factors in the scalar denominator
correspond to
\begin{align}
E_{f,0}^{(s_f)}
&=
\sqrt{M^2+(p_3-s_f\mu_5)^2},
\nonumber\\
E_{f,0}^{(-s_f)}
&=
\sqrt{M^2+(p_3+s_f\mu_5)^2}.
\label{eq:app-pi0-LLL-energies}
\end{align}
where
$s_f=\operatorname{sgn}(q_fB)$.
Accordingly,
\begin{equation}
U_{f,0}(p_0)
=
\left[p_0^2-(E_{f,0}^{(s_f)})^2\right]
\left[p_0^2-(E_{f,0}^{(-s_f)})^2\right].
\end{equation}
The corresponding numerator is
\begin{align}
h_{f;00}
={}&
-4\pi B_f
\left[
(E_{f,0}^{(s_f)})^2-p_0(p_0+K_0)
\right]
\nonumber\\
&\times
\left[p_0^2-(E_{f,0}^{(-s_f)})^2\right]
\nonumber\\
&\times
\left[(p_0+K_0)^2
-(E_{f,0}^{(-s_f)})^2\right].
\label{eq:app-pi0-h-LLL}
\end{align}
After canceling the opposite spin-state factors,
we obtain
\begin{align}
&\frac{h_{f;00}}
{U_{f,0}(p_0+K_0)U_{f,0}(p_0)}
\nonumber\\
={}&
-4\pi B_f
\frac{
(E_{f,0}^{(s_f)})^2-p_0(p_0+K_0)
}{
\left[p_0^2-(E_{f,0}^{(s_f)})^2\right]
\left[(p_0+K_0)^2
-(E_{f,0}^{(s_f)})^2\right]
}.
\label{eq:app-pi0-LLL-same-branch}
\end{align}

The remaining fermionic Matsubara sum can be
performed for each surviving energy branch:
\begin{align}
&T\sum_{\ell}
\frac{
E^2-p_0(p_0+K_0)
}{
(p_0^2-E^2)
[(p_0+K_0)^2-E^2]
}
\nonumber\\
&\qquad=
\frac{
2E\tanh(E/2T)
}{
4E^2-K_0^2
}.
\label{eq:app-pi0-matsubara}
\end{align}
Using this result, the polarization function
before folding the longitudinal momentum integral
can be expressed as
\begin{align}
\Pi_{\pi^0}(K_0)
={}&
\frac{N_c}{2\pi^2}
\sum_{f=u,d}B_f
\bigg[
\int_{-\infty}^{\infty}\dd p_3\,
\frac{
2E_{f,0}^{(s_f)}
\tanh(E_{f,0}^{(s_f)}/2T)
}{
4(E_{f,0}^{(s_f)})^2-K_0^2
}
\nonumber\\
&+
\sum_{n=1}^{\infty}\sum_{\eta=\pm1}
\int_{-\infty}^{\infty}\dd p_3\,
\frac{
2E_{f,n,\eta}
\tanh(E_{f,n,\eta}/2T)
}{
4E_{f,n,\eta}^2-K_0^2
}
\bigg].
\label{eq:app-pi0-unfolded}
\end{align}

For the HLL, the energies are even functions of
$p_3$, so that
\begin{equation}
\int_{-\infty}^{\infty}\dd p_3\,
F(E_{f,n,\eta})
=
2\int_0^\infty\dd p_3\,
F(E_{f,n,\eta}),
\qquad n\geq1.
\end{equation}
For the LLL, only one spin state is present at
fixed signed $p_3$. Folding the longitudinal
momentum integral instead gives
\begin{align}
&\int_{-\infty}^{\infty}\dd p_3\,
F(E_{f,0}^{(s_f)})
\nonumber\\
&\qquad=
\int_0^\infty\dd p_3
\left[
F(E_{f,0}^{(s_f)})
+
F(E_{f,0}^{(-s_f)})
\right].
\end{align}
The two folded LLL energies can be written as
\begin{equation}
E_{f,0,\eta}
=
\sqrt{M^2+(p_3+\eta\mu_5)^2},
\qquad \eta=\pm1,
\end{equation}
as defined in Eq.~\eqref{eq:quark-branches}.
Thus, the two $\eta$ contributions in the folded
LLL expression represent the positive- and
negative-$p_3$ sectors of the same spin state.

Combining the folded LLL and HLL contributions,
we obtain
\begin{align}
\Pi_{\pi^0}(K_0)
={}&
\frac{N_c}{4\pi^2}
\sum_{f=u,d}B_f
\int_0^\infty\dd p_3
\nonumber\\
&\bigg[
\sum_{\eta=\pm1}
\frac{
4E_{f,0,\eta}
\tanh(E_{f,0,\eta}/2T)
}{
4E_{f,0,\eta}^2-K_0^2
}
\nonumber\\
&+
2\sum_{n=1}^{\infty}\sum_{\eta=\pm1}
\frac{
4E_{f,n,\eta}
\tanh(E_{f,n,\eta}/2T)
}{
4E_{f,n,\eta}^2-K_0^2
}
\bigg]
\nonumber\\
={}&
\frac{N_c}{4\pi^2}
\sum_{f=u,d}B_f
\sum_{n=0}^{\infty}(2-\delta_{n0})
\sum_{\eta=\pm1}
\nonumber\\
&\times
\int_0^\infty\dd p_3\,
\frac{
4E_{f,n,\eta}
\tanh(E_{f,n,\eta}/2T)
}{
4E_{f,n,\eta}^2-K_0^2
}.
\label{eq:app-pi0-final}
\end{align}
The factor $2-\delta_{n0}$ accounts for the
different longitudinal-momentum folding factors
of the HLL and LLL, respectively.

The retarded polarization function follows from
the analytic continuation
\begin{equation}
K_0\longrightarrow\omega+i0^+.
\end{equation}
Taking its real part gives
\begin{align}
\operatorname{Re}\Pi_{\pi^0}^{R}(\omega)
={}&
\frac{N_c}{4\pi^2}
\sum_{f=u,d}B_f
\sum_{n=0}^{\infty}(2-\delta_{n0})
\sum_{\eta=\pm1}
\nonumber\\
&\times
\mathcal P\int_0^\infty\dd p_3\,
\frac{
4E_{f,n,\eta}
\tanh(E_{f,n,\eta}/2T)
}{
4E_{f,n,\eta}^2-\omega^2
}.
\label{eq:app-pi0-retarded}
\end{align}

The final expression contains only same-branch
pair-production denominators. Consequently, no
finite-energy mixed-branch thermal-scattering
contribution survives at zero external spatial
momentum.

\section{Explicit reduction of the charged-pion residue kernels}
\label{app:charged-reduction}

In this appendix, we present the explicit reduction of
the positive-energy residue kernels used in the numerical
calculation of the charged-pion channel.

After folding the longitudinal momentum integral onto
$p_3\geq0$, we define, for fixed Landau levels $(n,n')$,
\begin{align}
  P&=2n'B_u,
  &Q&=2nB_d,
  &r&=p_3^2,
  \nonumber\\
  \rho_u&=\sqrt{P+r},
  &\rho_d&=\sqrt{Q+r},
  \label{eq:appendix-short-definitions}
\end{align}
together with the corresponding quark energies
\begin{align}
  E_{u,\eta}
  &=\sqrt{M^2+(\rho_u+\eta\mu_5)^2},
  \nonumber\\
  E_{d,\eta}
  &=\sqrt{M^2+(\rho_d+\eta\mu_5)^2}.
  \label{eq:appendix-energies}
\end{align}

According to the pole ordering adopted below,
the two longitudinal kernels in
Eq.~\eqref{eq:charged-two-overlaps} can be written as
\begin{equation}
  \mathcal K_{nn'}^{(a)}
  =
  \sum_{i=1}^{8}T_{ai},
  \qquad a=1,2.
  \label{eq:appendix-kernel-map}
\end{equation}
Here, $a=1$ and $a=2$ correspond to the longitudinal
kernels multiplying $j_{nn'}^{++}$ and $j_{nn'}^{--}$,
respectively, in Eq.~\eqref{eq:charged-two-overlaps}.
Their relative sign is already included in that equation.

The four positive-energy poles are ordered as
\begin{equation}
  (E_4,E_3,E_2,E_1)
  =
  (E_{u,+},E_{u,-},E_{d,+},E_{d,-}).
  \label{eq:appendix-pole-ordering}
\end{equation}
Accordingly, $T_{a,2m}$ denotes the positive-energy
residue associated with $E_m$, whereas
$T_{a,2m-1}$ denotes the corresponding
negative-energy residue, with $m=1,\ldots,4$.

After taking the residues at the positive-energy poles,
the denominator of the remaining internal quark
propagator is given by
\begin{align}
  \Delta_{u,\eta}(\omega)
  &=
  \prod_{\eta'=\pm1}
  \left[
  (\omega-E_{u,\eta})^2
  -(E_{d,\eta'})^2
  \right],
  \label{eq:appendix-delta-u}
  \\
  \Delta_{d,\eta}(\omega)
  &=
  \prod_{\eta'=\pm1}
  \left[
  (\omega+E_{d,\eta})^2
  -(E_{u,\eta'})^2
  \right].
  \label{eq:appendix-delta-d}
\end{align}
Here, $\Delta_{u,\eta}$ and $\Delta_{d,\eta}$
denote the remaining denominator factors evaluated
at the positive-energy poles of the $u$ and $d$
quarks, respectively. They do not include the
pole-normalization factors of the propagators
that generate the corresponding poles.

The zeros of these denominators determine the
singularities associated with pair production
and thermal scattering. For positive external energy,
the corresponding conditions are
\begin{equation}
  \omega=E_{u,\eta}+E_{d,\eta'},
  \qquad
  \omega=
  \left|E_{u,\eta}-E_{d,\eta'}\right|.
\end{equation}
These correspond to the pair-production and
thermal-scattering processes discussed in
Eqs.~\eqref{eq:charged-unitary-process}
and \eqref{eq:charged-landau-process}, respectively.

We also define
\begin{equation}
  A_u=\omega^2+P-Q,
  \qquad
  A_d=\omega^2-P+Q.
  \label{eq:appendix-Aud}
\end{equation}

The on-shell numerator functions associated with
the first transverse-overlap structure are
\begin{align}
H_{u,\eta}^{(1)}
={}&
\frac{\eta\mu_5}{\rho_u}
\Big[
  \omega^2(5P+4r)
  +(P-Q)(3P+2r)
\nonumber\\
&\qquad
  -4\omega E_{u,\eta}(P+r)
\Big]
\nonumber\\
&+
\omega^2(2M^2+3P+2r+2\mu_5^2)
\nonumber\\
&+
(P-Q)(P+2\mu_5^2)
\nonumber\\
&-
\omega E_{u,\eta}(\omega^2+3P-Q),
\label{eq:appendix-h1u}
\\[3pt]
H_{d,\eta}^{(1)}
={}&
\frac{\eta\mu_5}{\rho_d}
\Big[
  \omega^2(5Q+4r)
  +(Q-P)(3Q+2r)
\nonumber\\
&\qquad
  +4\omega E_{d,\eta}(Q+r)
\Big]
\nonumber\\
&+
\omega^2(2M^2+3Q+2r+2\mu_5^2)
\nonumber\\
&+
(Q-P)(Q+2\mu_5^2)
\nonumber\\
&+
\omega E_{d,\eta}(\omega^2-P+3Q).
\label{eq:appendix-h1d}
\end{align}

For fixed positive $\mu_5$, the four positive-energy
residue kernels associated with the first overlap
structure are
\begin{align}
  T_{18}
  &=
  -\frac{n_F(E_{u,+})H_{u,+}^{(1)}}
  {4E_{u,+}\Delta_{u,+}},
  \nonumber\\
  T_{16}
  &=
  -\frac{n_F(E_{u,-})H_{u,-}^{(1)}}
  {4E_{u,-}\Delta_{u,-}},
  \nonumber\\
  T_{14}
  &=
  -\frac{n_F(E_{d,+})H_{d,+}^{(1)}}
  {4E_{d,+}\Delta_{d,+}},
  \nonumber\\
  T_{12}
  &=
  -\frac{n_F(E_{d,-})H_{d,-}^{(1)}}
  {4E_{d,-}\Delta_{d,-}}.
  \label{eq:appendix-four-t1}
\end{align}

The numerator functions associated with the second
transverse-overlap structure are
\begin{align}
  H_{u,\eta}^{(2)}
  ={}&
  \sqrt{PQ}
  \left[
  \left(1+\frac{\eta\mu_5}{\rho_u}\right)A_u
  -2\omega E_{u,\eta}
  \right],
  \nonumber\\
  H_{d,\eta}^{(2)}
  ={}&
  \sqrt{PQ}
  \left[
  \left(1+\frac{\eta\mu_5}{\rho_d}\right)A_d
  +2\omega E_{d,\eta}
  \right].
  \label{eq:appendix-h2ud}
\end{align}

The corresponding four positive-energy residue
kernels are
\begin{align}
  T_{28}
  &=
  \frac{n_F(E_{u,+})H_{u,+}^{(2)}}
  {4E_{u,+}\Delta_{u,+}},
  \nonumber\\
  T_{26}
  &=
  \frac{n_F(E_{u,-})H_{u,-}^{(2)}}
  {4E_{u,-}\Delta_{u,-}},
  \nonumber\\
  T_{24}
  &=
  \frac{n_F(E_{d,+})H_{d,+}^{(2)}}
  {4E_{d,+}\Delta_{d,+}},
  \nonumber\\
  T_{22}
  &=
  \frac{n_F(E_{d,-})H_{d,-}^{(2)}}
  {4E_{d,-}\Delta_{d,-}}.
  \label{eq:appendix-four-t2}
\end{align}

Here, $H_{u,\eta}^{(1,2)}$ and
$H_{d,\eta}^{(1,2)}$ are the reduced on-shell
numerator functions obtained from the residue
calculation, while $\Delta_{u,\eta}$ and
$\Delta_{d,\eta}$ are the corresponding remaining
denominators. The factors $4E_{f,\eta}$ are the
energy-normalization factors in the reduced residue
expressions, with the branch-dependent signs
included in the corresponding numerators and
residue expressions.

At nonzero temperature, the relation between
the Fermi distribution functions gives
the following connection between the positive-
and negative-energy residues for the four branches
defined in Eq.~\eqref{eq:appendix-pole-ordering}:
\begin{equation}
  T_{a,2m-1}(\omega)
  =
  -
  \frac{1-n_F(E_m)}{n_F(E_m)}
  T_{a,2m}(-\omega),
  \qquad
  a=1,2.
  \label{eq:appendix-negative-energy-partners}
\end{equation}

When performing a partial-fraction decomposition
of Eqs.~\eqref{eq:appendix-delta-u}
and \eqref{eq:appendix-delta-d},
inverse differences between the two energy branches
may appear. For example,
\begin{align}
  (E_{d,-})^2-(E_{d,+})^2
  &=
  (\rho_d-\mu_5)^2
  -(\rho_d+\mu_5)^2
  \nonumber\\
  &=
  -4\mu_5\rho_d.
\end{align}
Consequently,
\begin{equation}
  \left[
    (E_{d,-})^2-(E_{d,+})^2
  \right]^{-1}
  =
  -\frac{1}{4\mu_5\rho_d}.
  \label{eq:appendix-branch-degeneracy}
\end{equation}

As $\mu_5\to0$, the two energy branches become
degenerate, and the above coefficient exhibits
an apparent singularity. However, this singularity
arises from the partial-fraction decomposition
of the branch denominators, rather than from
a physical singularity of the original integration
kernel. The corresponding contributions cancel
when the complete branch sum is taken, ensuring
that the $\mu_5\to0$ limit remains regular.

\bibliography{references}

@article{Skokov2009Add,
  author = {V. Skokov and A. Yu. Illarionov and V. Toneev},
  title = {Estimate of the magnetic field strength in heavy-ion collisions},
  journal = {Int. J. Mod. Phys. A},
  volume = {24},
  pages = {5925--5932},
  year = {2009},
  doi = {10.1142/S0217751X09047570},
  eprint = {0907.1396},
  archivePrefix = {arXiv},
}

@article{Voronyuk2011Add,
  author = {V. Voronyuk and V. D. Toneev and W. Cassing and E. L. Bratkovskaya and V. P. Konchakovski and S. A. Voloshin},
  title = {Electromagnetic field evolution in relativistic heavy-ion collisions},
  journal = {Phys. Rev. C},
  volume = {83},
  pages = {054911},
  year = {2011},
  doi = {10.1103/PhysRevC.83.054911},
  eprint = {1103.4239},
  archivePrefix = {arXiv},
}

@article{DengHuang2012Add,
  author = {Wei-Tian Deng and Xu-Guang Huang},
  title = {Event-by-event generation of electromagnetic fields in heavy-ion collisions},
  journal = {Phys. Rev. C},
  volume = {85},
  pages = {044907},
  year = {2012},
  doi = {10.1103/PhysRevC.85.044907},
  eprint = {1201.5108},
  archivePrefix = {arXiv},
}

@article{Tuchin2016Add,
  author = {Kirill Tuchin},
  title = {Initial value problem for magnetic fields in heavy ion collisions},
  journal = {Phys. Rev. C},
  volume = {93},
  pages = {014905},
  year = {2016},
  doi = {10.1103/PhysRevC.93.014905},
  eprint = {1508.06925},
  archivePrefix = {arXiv},
}

@article{Gusynin1996,
  author  = {Gusynin, V. P. and Miransky, V. A. and Shovkovy, I. A.},
  title   = {Dimensional reduction and catalysis of dynamical symmetry breaking by a magnetic field},
  journal = {Nucl. Phys. B},
  volume  = {462},
  pages   = {249--290},
  year    = {1996},
  doi     = {10.1016/0550-3213(96)00021-1},
  eprint  = {hep-ph/9509320}
}

@article{Miransky2015,
  author = {Miransky, V. A. and Shovkovy, I. A.},
  title = {Quantum field theory in a magnetic field:
           From quantum chromodynamics to graphene
           and {Dirac} semimetals},
  journal = {Phys. Rept.},
  volume = {576},
  pages = {1--209},
  year = {2015},
  doi = {10.1016/j.physrep.2015.02.003},
  eprint = {1503.00732},
  archivePrefix = {arXiv},
  primaryClass = {hep-ph}
}

@article{Bali2012a,
  author  = {Bali, G. S. and Bruckmann, F. and Endrodi, G. and Fodor, Z. and Katz, S. D. and Krieg, S. and Schafer, A. and Szabo, K. K.},
  title   = {The {QCD} phase diagram for external magnetic fields},
  journal = {J. High Energy Phys.},
  volume  = {02},
  number  = {2012},
  pages   = {044},
  year    = {2012},
  doi     = {10.1007/JHEP02(2012)044},
  eprint  = {1111.4956},
  archivePrefix = {arXiv},
  primaryClass  = {hep-lat}
}

@article{Bali2012b,
  author = {Bali, G. S. and Bruckmann, F. and Endrodi, G.
            and Fodor, Z. and Katz, S. D. and Schafer, A.},
  title = {{QCD} quark condensate in external magnetic fields},
  journal = {Phys. Rev. D},
  volume = {86},
  pages = {071502},
  year = {2012},
  doi = {10.1103/PhysRevD.86.071502},
  eprint = {1206.4205},
  archivePrefix = {arXiv},
  primaryClass = {hep-lat}
}

@article{Coppola2018,
  author  = {Coppola, M. and Gomez Dumm, D. and Scoccola, N. N.},
  title   = {Charged pion masses under strong magnetic fields in the {NJL} model},
  journal = {Phys. Lett. B},
  volume  = {782},
  pages   = {155--161},
  year    = {2018},
  doi     = {10.1016/j.physletb.2018.04.043},
  eprint  = {1802.08041},
  archivePrefix = {arXiv},
  primaryClass  = {hep-ph}
}

@article{Coppola2019,
  author  = {Coppola, M. and Gomez Dumm, D. and Noguera, S. and Scoccola, N. N.},
  title   = {Neutral and charged pion properties under strong magnetic fields in the {NJL} model},
  journal = {Phys. Rev. D},
  volume  = {100},
  pages   = {054014},
  year    = {2019},
  doi     = {10.1103/PhysRevD.100.054014},
  eprint  = {1907.05840},
  archivePrefix = {arXiv},
  primaryClass  = {hep-ph}
}

@article{Bali2018Add,
  author = {Gunnar S. Bali and Bastian B. Brandt and Gergely Endrodi and Benjamin Glaessle},
  title = {Meson masses in electromagnetic fields with {Wilson} fermions},
  journal = {Phys. Rev. D},
  volume = {97},
  pages = {034505},
  year = {2018},
  doi = {10.1103/PhysRevD.97.034505},
  eprint = {1707.05600},
  archivePrefix = {arXiv},
}

@article{Ding2021,
  author  = {Ding, H.-T. and Li, S.-T. and Tomiya, A. and Wang, X.-D. and Zhang, Y.},
  title   = {Chiral properties of $(2+1)$-flavor {QCD} in strong magnetic fields at zero temperature},
  journal = {Phys. Rev. D},
  volume  = {104},
  pages   = {014505},
  year    = {2021},
  doi     = {10.1103/PhysRevD.104.014505},
  eprint  = {2008.00493},
  archivePrefix = {arXiv},
  primaryClass  = {hep-lat}
}

@article{Ding2026,
  author  = {Ding, Heng-Tong and Zhang, Dan},
  title   = {Chiral properties of $(2+1)$-flavor {QCD} in magnetic fields at zero temperature},
  journal = {Phys. Rev. D},
  volume  = {113},
  pages   = {094503},
  year    = {2026},
  doi     = {10.1103/qpxq-xjqc},
  eprint  = {2601.18354},
  archivePrefix = {arXiv},
  primaryClass  = {hep-lat}
}

@article{Lin:2022ied,
    author = "Lin, Fan and Xu, Kun and Huang, Mei",
    title = "{Magnetism of QCD matter and the pion mass from tensor-type spin polarization and the anomalous magnetic moment of quarks}",
    eprint = "2202.03226",
    archivePrefix = "arXiv",
    primaryClass = "hep-ph",
    doi = "10.1103/PhysRevD.106.016005",
    journal = "Phys. Rev. D",
    volume = "106",
    number = "1",
    pages = "016005",
    year = "2022"
}

@article{Xing:2021kbw,
    author = "Xing, Zanbin and Chao, Jingyi and Chang, Lei and Liu, Yu-xin",
    title = "{Exposing the effect of the p-wave component in the pion triplet under a strong magnetic field}",
    eprint = "2110.01245",
    archivePrefix = "arXiv",
    primaryClass = "hep-ph",
    doi = "10.1103/PhysRevD.105.114003",
    journal = "Phys. Rev. D",
    volume = "105",
    number = "11",
    pages = "114003",
    year = "2022"
}

@article{Zhai2026Review,
  author  = {Zhai, Shijie and Mei, Jie and Yu, Lang and Huang, Mei},
  title   = {Magnetized {QCD} Matter},
  journal = {Universe},
  volume  = {12},
  number  = {6},
  pages   = {154},
  year    = {2026},
  doi     = {10.3390/universe12060154}
}

@article{Mao2019,
  author  = {Mao, Shijun},
  title   = {Pions in magnetic field at finite temperature},
  journal = {Phys. Rev. D},
  volume  = {99},
  pages   = {056005},
  year    = {2019},
  doi     = {10.1103/PhysRevD.99.056005},
  eprint  = {1808.10242},
  archivePrefix = {arXiv},
  primaryClass  = {nucl-th}
}

@article{Sheng2021,
  author  = {Sheng, B. and Wang, Y. and Wang, X. and Yu, L.},
  title   = {Pole and screening masses of neutral pions in a hot and magnetized medium: A comprehensive study in the {Nambu--Jona-Lasinio} model},
  journal = {Phys. Rev. D},
  volume  = {103},
  pages   = {094001},
  year    = {2021},
  doi     = {10.1103/PhysRevD.103.094001},
  eprint  = {2010.05716},
  archivePrefix = {arXiv},
  primaryClass  = {hep-ph}
}

@article{LiMao2023Mott,
  author = {Li, Luyang and Mao, Shijun},
  title = {Inverse magnetic catalysis effect and current quark mass effect on mass spectra and Mott transitions of pions under external magnetic field},
  journal = {Phys. Rev. D},
  volume = {108},
  number = {5},
  pages = {054001},
  year = {2023},
  doi = {10.1103/PhysRevD.108.054001},
  eprint = {2308.12491},
  archivePrefix = {arXiv},
  primaryClass = {hep-ph}
}

@article{Mei2026,
  author  = {Mei, Jie and Wen, Rui and Zhou, Min and Mao, Shijun and Huang, Mei},
  title   = {Spectral function for pions in a magnetic field},
  journal = {Phys. Rev. D},
  volume  = {113},
  pages   = {074031},
  doi     = {10.1103/skmg-ql8c},
  year    = {2026},
  eprint  = {2601.22422},
  archivePrefix = {arXiv},
  primaryClass  = {hep-ph}
}

@article{MooreTassler2011Add,
  author = {Guy D. Moore and Marcus Tassler},
  title = {The sphaleron rate in {SU(N)} gauge theory},
  journal = {J. High Energy Phys.},
  volume  = {02},
  number  = {2011},
  pages   = {105},
  year    = {2011},
  doi = {10.1007/JHEP02(2011)105},
  eprint = {1011.1167},
  archivePrefix = {arXiv},
}

@article{Kharzeev2008,
  author  = {Kharzeev, D. E. and McLerran, L. D. and Warringa, H. J.},
  title   = {The effects of topological charge change in heavy ion collisions: Event by event {P} and {CP} violation},
  journal = {Nucl. Phys. A},
  volume  = {803},
  pages   = {227--253},
  year    = {2008},
  doi     = {10.1016/j.nuclphysa.2008.02.298},
  eprint  = {0711.0950},
  archivePrefix = {arXiv},
  primaryClass  = {hep-ph}
}

@article{Fukushima2008,
  author  = {Fukushima, K. and Kharzeev, D. E. and Warringa, H. J.},
  title   = {Chiral magnetic effect},
  journal = {Phys. Rev. D},
  volume  = {78},
  pages   = {074033},
  year    = {2008},
  doi     = {10.1103/PhysRevD.78.074033},
  eprint  = {0808.3382},
  archivePrefix = {arXiv},
  primaryClass  = {hep-ph}
}

@article{Braguta2016Add,
  author = {V. V. Braguta and E.-M. Ilgenfritz and A. Yu. Kotov and B. Petersson and S. A. Skinderev},
  title = {Study of the {QCD} phase diagram with a nonzero chiral chemical potential},
  journal = {Phys. Rev. D},
  volume = {93},
  pages = {034509},
  year = {2016},
  doi = {10.1103/PhysRevD.93.034509},
}

@article{Yu2016,
  author  = {Yu, L. and Liu, H. and Huang, M.},
  title   = {Effect of the chiral chemical potential on the chiral phase transition in the {NJL} model with different regularization schemes},
  journal = {Phys. Rev. D},
  volume  = {94},
  pages   = {014026},
  year    = {2016},
  doi     = {10.1103/PhysRevD.94.014026},
  eprint  = {1511.03073},
  archivePrefix = {arXiv},
  primaryClass  = {hep-ph}
}

@article{RuggieriPeng2016Add,
  author = {M. Ruggieri and G. X. Peng},
  title = {Critical temperature of chiral symmetry restoration for quark matter with a chiral chemical potential},
  journal = {J. Phys. G},
  volume = {43},
  pages = {125101},
  year = {2016},
  doi = {10.1088/0954-3899/43/12/125101},
  eprint = {1602.05250},
  archivePrefix = {arXiv},
}

@article{Ghosh2024,
  author  = {Ghosh, S. and Chaudhuri, N. and Sarkar, S. and Roy, P.},
  title   = {Mass and spectral function of scalar and pseudoscalar mesons in a hot and chirally imbalanced medium using the two-flavor {NJL} model},
  journal = {Phys. Rev. D},
  volume  = {109},
  pages   = {016021},
  year    = {2024},
  doi     = {10.1103/PhysRevD.109.016021},
  eprint  = {2312.08652},
  archivePrefix = {arXiv},
  primaryClass  = {hep-ph}
}

@article{Nambu1961a,
  author  = {Nambu, Y. and Jona-Lasinio, G.},
  title   = {Dynamical model of elementary particles based on an analogy with superconductivity. {I}},
  journal = {Phys. Rev.},
  volume  = {122},
  pages   = {345--358},
  year    = {1961},
  doi     = {10.1103/PhysRev.122.345}
}

@article{Nambu1961b,
  author  = {Nambu, Y. and Jona-Lasinio, G.},
  title   = {Dynamical model of elementary particles based on an analogy with superconductivity. {II}},
  journal = {Phys. Rev.},
  volume  = {124},
  pages   = {246--254},
  year    = {1961},
  doi     = {10.1103/PhysRev.124.246}
}

@article{Klevansky1992,
  author  = {Klevansky, S. P.},
  title   = {The {Nambu--Jona-Lasinio} model of quantum chromodynamics},
  journal = {Rev. Mod. Phys.},
  volume  = {64},
  pages   = {649--708},
  year    = {1992},
  doi     = {10.1103/RevModPhys.64.649}
}

@article{Hatsuda1994,
  author  = {Hatsuda, T. and Kunihiro, T.},
  title   = {{QCD} phenomenology based on a chiral effective {Lagrangian}},
  journal = {Phys. Rept.},
  volume  = {247},
  pages   = {221--367},
  year    = {1994},
  doi     = {10.1016/0370-1573(94)90022-1},
  eprint  = {hep-ph/9401310}
}

@article{Buballa2005,
  author  = {Buballa, M.},
  title = {{NJL}-model analysis of dense quark matter},
  journal = {Phys. Rept.},
  volume  = {407},
  pages   = {205--376},
  year    = {2005},
  doi     = {10.1016/j.physrep.2004.11.004},
  eprint  = {hep-ph/0402234}
}

@article{Yamamoto2011Add,
  author = {Arata Yamamoto},
  title = {Chiral magnetic effect in lattice {QCD} with a chiral chemical potential},
  journal = {Phys. Rev. Lett.},
  volume = {107},
  pages = {031601},
  year = {2011},
  doi = {10.1103/PhysRevLett.107.031601},
  eprint = {1105.0385},
  archivePrefix = {arXiv},
}

@article{Chaudhuri2022,
  author = {Chaudhuri, Nilanjan and Mukherjee, Arghya
            and Ghosh, Snigdha and Sarkar, Sourav and Roy, Pradip},
  title = {Insignificance of the anomalous magnetic moment
           of the quarks in presence of chiral imbalance},
  journal = {Eur. Phys. J. A},
  volume = {58},
  pages = {82},
  year = {2022},
  doi = {10.1140/epja/s10050-022-00731-9},
  eprint = {2111.12058},
  archivePrefix = {arXiv},
  primaryClass = {hep-ph}
}

@article{Schwinger1951,
  author  = {Schwinger, J. S.},
  title   = {On gauge invariance and vacuum polarization},
  journal = {Phys. Rev.},
  volume  = {82},
  pages   = {664--679},
  year    = {1951},
  doi     = {10.1103/PhysRev.82.664}
}

@article{Ritus1972,
  author  = {Ritus, V. I.},
  title   = {Radiative corrections in quantum electrodynamics with intense field and their analytical properties},
  journal = {Ann. Phys.},
  volume  = {69},
  pages   = {555--582},
  year    = {1972},
  doi     = {10.1016/0003-4916(72)90191-1}
}

@article{PauliVillars1949,
  author  = {Pauli, W. and Villars, F.},
  title   = {On the invariant regularization in relativistic quantum theory},
  journal = {Rev. Mod. Phys.},
  volume  = {21},
  pages   = {434--444},
  year    = {1949},
  doi     = {10.1103/RevModPhys.21.434}
}

@article{Avancini2017,
  author  = {Avancini, Sidney S. and Farias, Ricardo L. S. and Pinto, Marcus Benghi and Tavares, William R. and Timoteo, Varese S.},
  title   = {$\pi_0$ pole mass calculation in a strong magnetic field and lattice constraints},
  journal = {Phys. Lett. B},
  volume  = {767},
  pages   = {247--252},
  year    = {2017},
  doi     = {10.1016/j.physletb.2017.02.002},
  eprint  = {1606.05754},
  archivePrefix = {arXiv},
  primaryClass  = {hep-ph}
}

@article{Hattori2016,
  author  = {Hattori, Koichi and Kojo, Toru and Su, Nan},
  title   = {Mesons in strong magnetic fields: ({I}) General analyses},
  journal = {Nucl. Phys. A},
  volume  = {951},
  pages   = {1--30},
  year    = {2016},
  doi     = {10.1016/j.nuclphysa.2016.03.016},
  eprint  = {1512.07361},
  archivePrefix = {arXiv},
  primaryClass  = {hep-ph}
}

@article{Andreichikov2017,
  author  = {Andreichikov, M. A. and Kerbikov, B. O. and Luschevskaya, E. V. and Simonov, Yu. A. and Solovjeva, O. E.},
  title   = {The evolution of meson masses in a strong magnetic field},
  journal = {J. High Energy Phys.},
  volume  = {05},
  number  = {2017},
  pages   = {007},
  year    = {2017},
  doi     = {10.1007/JHEP05(2017)007},
  eprint  = {1610.06887},
  archivePrefix = {arXiv},
  primaryClass  = {hep-ph}
}

@article{HattoriYamamoto2019,
    author = {Hattori, Koichi and Yamamoto, Arata},
    title = {Meson deformation by magnetic fields in lattice QCD},
    eprint = {1901.10182},
    archivePrefix = {arXiv},
    primaryClass = {hep-lat},
    doi = {10.1093/ptep/ptz023},
    journal = {Prog. Theor. Exp. Phys.},
    volume = {2019},
    number = {4},
    pages = {043B04},
    year = {2019}
}

@article{Andersen2012Add,
    author = {Andersen, Jens O.},
    title = {Thermal pions in a magnetic background},
    eprint = {1202.2051},
    archivePrefix = {arXiv},
    primaryClass = {hep-ph},
    doi = {10.1103/PhysRevD.86.025020},
    journal = {Phys. Rev. D},
    volume = {86},
    number = {2},
    pages = {025020},
    year = {2012}
}

@article{Luschevskaya2016Add,
  author = {E. V. Luschevskaya and O. E. Solovjeva and O. V. Teryaev},
  title = {Magnetic polarizability of pion},
  journal = {Phys. Lett. B},
  volume = {761},
  pages = {393--398},
  year = {2016},
  doi = {10.1016/j.physletb.2016.08.054},
  eprint = {1511.09316},
  archivePrefix = {arXiv},
}

@article{WangMixing2026Add,
  author = {Ziyue Wang},
  title = {Comparative study of mass extraction schemes and $\pi^{\pm}$--$\rho^{\pm}$ mixing},
  journal = {Phys. Rev. D},
  volume = {114},
  number = {7},
  pages = {074036},
  year = {2026},
  doi = {10.1103/62jw-g457},
  eprint = {2605.06271},
  archivePrefix = {arXiv},
  primaryClass = {hep-ph}
}

@article{Wen:2023qcz,
    author = {Wen, Rui and Yin, Shi and Fu, Wei-jie and Huang, Mei},
    title = {Functional renormalization group study of neutral
             and charged pions in magnetic fields in the
             quark-meson model},
    journal = {Phys. Rev. D},
    volume = {108},
    number = {7},
    pages = {076020},
    year = {2023},
    doi = {10.1103/PhysRevD.108.076020},
    eprint = {2306.04045},
    archivePrefix = {arXiv},
    primaryClass = {hep-ph}
}

@article{GomezDumm:2023owj,
    author = {G{\'o}mez Dumm, D. and Noguera, S. and Scoccola, N. N.},
    title = {Charged meson masses under strong magnetic fields:
             Gauge invariance and {Schwinger} phases},
    journal = {Phys. Rev. D},
    volume = {108},
    number = {1},
    pages = {016012},
    year = {2023},
    doi = {10.1103/PhysRevD.108.016012},
    eprint = {2306.04128},
    archivePrefix = {arXiv},
    primaryClass = {hep-ph}
}

@article{Chao:2013qpa,
    author = {Chao, Jingyi and Chu, Pengcheng and Huang, Mei},
    title = {Inverse magnetic catalysis induced by sphalerons},
    journal = {Phys. Rev. D},
    volume = {88},
    pages = {054009},
    year = {2013},
    doi = {10.1103/PhysRevD.88.054009},
    eprint = {1305.1100},
    archivePrefix = {arXiv},
    primaryClass = {hep-ph}
}

@article{Fayazbakhsh:2012vr,
  author = {Fayazbakhsh, Sh. and Sadeghian, S. and Sadooghi, N.},
  title = {Properties of neutral mesons in a hot and magnetized quark matter},
  journal = {Phys. Rev. D},
  volume = {86},
  number = {8},
  pages = {085042},
  year = {2012},
  doi = {10.1103/PhysRevD.86.085042},
  eprint = {1206.6051},
  archivePrefix = {arXiv},
  primaryClass = {hep-ph}
}

@article{Avancini:2018svs,
  author = {Avancini, Sidney S. and Farias, Ricardo L. S.
            and Tavares, William R.},
  title = {Neutral meson properties in hot and magnetized
         quark matter: A new magnetic field independent
         regularization scheme applied to an NJL-type model},
  journal = {Phys. Rev. D},
  volume = {99},
  number = {5},
  pages = {056009},
  year = {2019},
  doi = {10.1103/PhysRevD.99.056009},
  eprint = {1812.00945},
  archivePrefix = {arXiv},
  primaryClass = {hep-ph}
}

@article{Liu:2018zag,
  author = {Liu, Hao and Wang, Xinyang and Yu, Lang
            and Huang, Mei},
  title = {Neutral and charged scalar mesons,
           pseudoscalar mesons, and diquarks in magnetic fields},
  journal = {Phys. Rev. D},
  volume = {97},
  number = {7},
  pages = {076008},
  year = {2018},
  doi = {10.1103/PhysRevD.97.076008}
}

@article{LiWang2020,
  author  = {Li, Wei and Wang, Gang},
  title   = {Chiral Magnetic Effects in Nuclear Collisions},
  journal = {Annu. Rev. Nucl. Part. Sci.},
  volume  = {70},
  pages   = {293--321},
  year    = {2020},
  doi     = {10.1146/annurev-nucl-030220-065203}
}

@article{Astrakhantsev2021,
  author = {Astrakhantsev, Nikita and Braguta, V. V.
            and Kotov, A. Yu. and Kuznedelev, D. D.
            and Nikolaev, A. A.},
  title = {Lattice study of QCD at finite chiral density:
           topology and confinement},
  journal = {Eur. Phys. J. A},
  volume = {57},
  pages = {15},
  year = {2021},
  doi = {10.1140/epja/s10050-020-00326-2}
}

@article{Ruggieri2020,
  author = {Ruggieri, Marco and
            Chernodub, Maxim N. and
            Lu, Zhen-Yan},
  title = {Topological susceptibility, divergent chiral
           density, and phase diagram of chirally
           imbalanced QCD medium at finite temperature},
  journal = {Phys. Rev. D},
  volume = {102},
  pages = {014031},
  year = {2020},
  doi = {10.1103/PhysRevD.102.014031},
  eprint = {2004.09393},
  archivePrefix = {arXiv}
}

@article{Espriu2020,
  author = {Espriu, D. and
            G{\'o}mez Nicola, A. and
            Vioque-Rodr{\'i}guez, A.},
  title = {Chiral perturbation theory for nonzero
           chiral imbalance},
  journal = {J. High Energy Phys.},
  volume = {06},
  number = {2020},
  pages = {062},
  year = {2020},
  doi = {10.1007/JHEP06(2020)062},
  eprint = {2002.11696},
  archivePrefix = {arXiv}
}

@article{BruckmannEndrodiKovacs2013,
  author  = {Bruckmann, Falk and Endrodi, Gergely and Kovacs, Tamas G.},
  title   = {Inverse magnetic catalysis and the Polyakov loop},
  journal = {Journal of High Energy Physics},
  volume  = {2013},
  number  = {04},
  pages   = {112},
  year    = {2013},
  doi     = {10.1007/JHEP04(2013)112}
}

@article{AndersenNaylorTranberg2016,
  author  = {Andersen, Jens O. and Naylor, William R. and Tranberg, Anders},
  title   = {Phase diagram of QCD in a magnetic field: A review},
  journal = {Reviews of Modern Physics},
  volume  = {88},
  pages   = {025001},
  year    = {2016},
  doi     = {10.1103/RevModPhys.88.025001}
}

@article{MuellerPawlowski2015,
  author  = {Mueller, Niklas and Pawlowski, Jan M.},
  title   = {Magnetic catalysis and inverse magnetic catalysis in QCD},
  journal = {Physical Review D},
  volume  = {91},
  pages   = {116010},
  year    = {2015},
  doi     = {10.1103/PhysRevD.91.116010}
}

@article{LiCaoHe2021,
  author  = {Li, Jianing and Cao, Gaoqing and He, Lianyi},
  title   = {Gauge independence of pion masses in a magnetic field within the Nambu--Jona-Lasinio model},
  journal = {Physical Review D},
  volume  = {104},
  pages   = {074026},
  year    = {2021},
  doi     = {10.1103/PhysRevD.104.074026}
}

@article{Coppola:2024mixing,
    author = "Coppola, M\'aximo and G\'omez Dumm, Daniel
              and Noguera, Santiago and Scoccola, Norberto N.",
    title = "{Masses of magnetized pseudoscalar and vector
             mesons in an extended NJL model:
             The role of axial vector mesons}",
    eprint = "2312.16675",
    archivePrefix = "arXiv",
    primaryClass = "hep-ph",
    doi = "10.1103/PhysRevD.109.054014",
    journal = "Phys. Rev. D",
    volume = "109",
    number = "5",
    pages = "054014",
    year = "2024"
}

@article{Huang:2015oca,
  author = {X.-G. Huang},
  title = {Electromagnetic fields and anomalous transports
           in heavy-ion collisions -- A pedagogical review},
  journal = {Rep. Prog. Phys.},
  volume = {79},
  number = {7},
  pages = {076302},
  year = {2016},
  doi = {10.1088/0034-4885/79/7/076302},
  eprint = {1509.04073},
  archivePrefix = {arXiv},
  primaryClass = {nucl-th}
}

@article{Wang:2017vtn,
  author = {Wang, Ziyue and Zhuang, Pengfei},
  title = {Meson properties in magnetized quark matter},
  journal = {Phys. Rev. D},
  volume = {97},
  number = {3},
  pages = {034026},
  year = {2018},
  doi = {10.1103/PhysRevD.97.034026},
  eprint = {1712.00554},
  archivePrefix = {arXiv},
  primaryClass = {hep-ph}
}

@article{Avancini:2015ady,
  author = {Avancini, Sidney S. and Tavares, William R. and Pinto, Marcus B.},
  title = {Properties of magnetized neutral mesons within a full RPA evaluation},
  journal = {Phys. Rev. D},
  volume = {93},
  number = {1},
  pages = {014010},
  year = {2016},
  doi = {10.1103/PhysRevD.93.014010},
  eprint = {1511.06261},
  archivePrefix = {arXiv},
  primaryClass = {hep-ph}
}

@article{Avancini:2021pmi,
  author = {Avancini, Sidney S. and Sodr{\'e}, Joana C. and Coppola, M{\'a}ximo and Scoccola, Norberto N.},
  title = {Light pseudoscalar meson masses under strong magnetic fields within the {SU(3)} {Nambu--Jona-Lasinio} model},
  journal = {Phys. Rev. D},
  volume = {104},
  number = {9},
  pages = {094040},
  year = {2021},
  doi = {10.1103/PhysRevD.104.094040},
  eprint = {2109.01911},
  archivePrefix = {arXiv},
  primaryClass = {hep-ph}
}

@article{Sheng:2021evj,
  author = {Sheng, Bing-kai and Wang, Xinyang and Yu, Lang},
  title = {Impacts of inverse magnetic catalysis on screening masses of neutral pions and sigma mesons in hot and magnetized quark matter},
  journal = {Phys. Rev. D},
  volume = {105},
  number = {3},
  pages = {034003},
  year = {2022},
  doi = {10.1103/PhysRevD.105.034003},
  eprint = {2110.12811},
  archivePrefix = {arXiv},
  primaryClass = {hep-ph}
}

@article{Mei:2022dkd,
  author = {Mei, Jie and Xia, Tao and Mao, Shijun},
  title = {Mass spectra of neutral mesons $K_0$, $\pi_0$, $\eta$, and $\eta^\prime$ at finite magnetic field, temperature and quark chemical potential},
  journal = {Phys. Rev. D},
  volume = {107},
  number = {7},
  pages = {074018},
  year = {2023},
  doi = {10.1103/PhysRevD.107.074018},
  eprint = {2212.04778},
  archivePrefix = {arXiv},
  primaryClass = {hep-ph},
  note = {Erratum: Phys. Rev. D 110, 119901(E) (2024)}
}

@article{Ding:2020jui,
  author = {Ding, Heng-Tong and Li, Sheng-Tai and Mukherjee, Swagato and Tomiya, Akio and Wang, Xiao-Dan},
  title = {Meson masses in external magnetic fields with {HISQ} fermions},
  journal = {PoS},
  volume = {LATTICE2019},
  pages = {250},
  year = {2020},
  doi = {10.22323/1.363.0250},
  eprint = {2001.05322},
  archivePrefix = {arXiv},
  primaryClass = {hep-lat}
}

@article{Xu:2020yag,
  author = {Xu, Kun and Chao, Jingyi and Huang, Mei},
  title = {Effect of the anomalous magnetic moment of quarks on magnetized {QCD} matter and meson spectra},
  journal = {Phys. Rev. D},
  volume = {103},
  number = {7},
  pages = {076015},
  year = {2021},
  doi = {10.1103/PhysRevD.103.076015},
  eprint = {2007.13122},
  archivePrefix = {arXiv},
  primaryClass = {hep-ph}
}

\end{document}